\documentclass[12pt, preprint]{aastex}

\begin{document}
\shorttitle{Evolution of Debris Disks}
\shortauthors{Currie, T. et al.}
\title{The Rise and Fall of Debris Disks: MIPS Observations of h and $\chi$ Persei and the Evolution of Mid-IR 
Emission from Planet Formation}
\author{Thayne Currie\altaffilmark{1,5}, Scott J. Kenyon\altaffilmark{1,6}, 
Zoltan Balog\altaffilmark{2,7},George Rieke\altaffilmark{2}, Ann Bragg\altaffilmark{3,6}, \&
Benjamin Bromley\altaffilmark{4}}
\altaffiltext{1}{Harvard-Smithsonian Center for Astrophysics, 60 Garden St. Cambridge, MA 02140}
\altaffiltext{2}{Steward Observatory, University of Arizona,  933 N. Cherry Av. Tucson, AZ 85721}
\altaffiltext{3}{Department of Physics, Bowling Green State University, Bowling Green, OH}
\altaffiltext{4}{Department of Physics, University of Utah, 201 JFB, Salt Lake City, UT 84112}
\altaffiltext{5}{Department of Physics \& Astronomy, University of California-Los Angeles, Los Angeles, 
CA, 90095}
\altaffiltext{6}{Visiting Astronomer, Kitt Peak National Observatory, National Optical Astronomy
Observatory, which is operated by the Association of Universities for Research
in Astronomy, Inc., under cooperative agreement with the National Science
Foundation.}
\altaffiltext{7}{on leave from Dept. of Optics and Quantum Electronics, University of Szeged, H-6720, Szeged, Hungary}
\email{tcurrie@cfa.harvard.edu}
\begin{abstract}
We describe Spitzer/MIPS observations of the double cluster, h and $\chi$ Persei, 
covering a $\sim$ 0.6 square-degree area surrounding the cores of both clusters.  
The data are combined with IRAC and 2MASS data to investigate $\sim$ 616 sources 
from 1.25-24 $\mu m$.  We use the long-baseline 
$K_{s}$-[24] color to identify two populations with IR excess indicative of circumstellar 
material: Be stars with 24 $\mu m$ excess from 
optically-thin free free emission and 17 fainter sources (J$\sim$ 14-15)
 with [24] excess consistent with a circumstellar disk.
The frequency of IR excess for the fainter sources increases 
from 4.5 $\mu m$ through 24 $\mu m$.  The IR excess is likely due to debris from the planet 
formation process.  The wavelength-dependent behavior is consistent with an inside-out clearing of circumstellar disks.
A comparison of the 24 $\mu m$ excess population in h and $\chi$ Per sources with results for 
 other clusters shows that 24 $\mu m$ emission from debris disks 'rises' from 5 to 10 Myr, peaks 
at $\sim$ 10-15 Myr, and then 'falls' from $\sim$ 15/20 Myr to 1 Gyr.
\end{abstract}
\section{Introduction}
Most 1-2 Myr-old stars are surrounded by massive (M$_{disk}$ $\sim$ 0.01-0.1 M$_{\star}$) 
optically-thick accretion disks of gas and dust.  The disk
 produces near-to-mid infrared (IR) emission comparable in brightness to the stellar photosphere (L$_{disk}$ 
$\sim$ L$_{\star}$) (e.g. Kenyon \& Hartmann 1995, Hillenbrand 1997).  
The evolution of these 'primordial' disks has been studied extensively (e.g. Haisch, Lada, \& Lada 2001; 
Lada et al. 2006; Dahm \& Hillenbrand 2007).  
By 5-10 Myr, primordial disks disappear and less massive (M$_{disk}$ $\lesssim$ 1 M$_{\oplus}$) 
gas-poor, optically-thin 'debris disks' 
with weaker emission (L$_{disk}$ $\lesssim$ 10$^{-3}$ L$_{\star}$) emerge 
(e.g. Hernandez et al. 2006).  By $\sim$ 10-20 Myr, primordial disks are extremely
 rare: almost all disks are debris disks (Currie et al. 2007a, hereafter C07a; Gorlova et al. 2007; 
Sicilia-Aguilar et al. 2006).  

Debris disks older than $\sim$ 20Myr are well studied. Rieke et al. (2005; hereafter
R05) showed that the 24 $\mu m$ emission declines with time as t$^{-1}$ (see also 
Kalas 1998; Habing et al. 2001; Decin et al. 2003).  This decay
agrees with expectations for the gradual depletion of the reservoir of small planetesimals. With fewer
parent bodies to initiate the collisional cascades that yield the infrared-emitting dust, the infrared excesses
drop systematically with time (Kenyon \& Bromley 2002; Dominik \& Decin 2003; R05; Wyatt et al. 2007a).  R05 also found a large range in the
amount of infrared excess emission at each age, even for very young systems. Wyatt et al. (2007a) demonstrate that the
first-order cause of this range is probably the large variation in protostellar disk masses and hence in the
mass available to form planetesimals.

Because the R05 sample and other studies of individual stars (e.g.,
Chen et al. 2005a) include few stars younger than 20Myr, they do not probe the 5-20 Myr
transitional period from primordial to debris disks well. This transition marks an important phase for planet
formation and other physical processes in disks.  
Gas accretion onto most young stars ceases by $\approx$ 10 Myr (Sicilia-Aguilar 
et al. 2005).  Planets acquire most of their mass
by $\approx$ 5-20 Myr (Kenyon \& Bromley 2006; Chambers 2001; Wetherill \& Stewart 1993).  

With an age of 13 $\pm$ 1 Myr and with over $\sim$ 5000 members (C07a), the double cluster, h and $\chi$ Persei 
(d=2.34 kpc, A$_{V}$$\sim$ 1.62; Slesnick et al. 2002, Bragg \& Kenyon 
2005), provides an ideal laboratory to study disk evolution during this critical age.  
Recent observations of h and $\chi$ Per with the 
Spitzer Space Telescope have demonstrated the utility of using the double cluster to investigate 
disk evolution after the primordial stage.  C07a used 3.6-8$\mu m$ Spitzer data to show that 
disks last longer around less massive stars and 
 at greater distances from the star.  C07b analyzed a well-constrained subsample of h and $\chi$ Per 
sources and showed that at least some of the disk emission in them comes from warm dust 
in the terrestrial zones of disks as a byproduct of terrestrial planet formation.  

In this paper, we use data obtained with the Multiband Imaging Photometer for Spitzer (MIPS) to extend
the study of h and $\chi$ Per to 24$\mu$m. This band allows us to search for high levels of mid-IR excess
associated with cool dust that orbits in a disk at $\sim$ 2-50 AU from the central star. Our survey
covers a region containing $\sim$ 600 intermediate-to-high mass cluster members.
In \S 2 we describe the MIPS observations, data reduction, and sample selection.  
We analyze the 24$\mu$m photometry in \S 3. The main results are: 1) there are two IR-excess
populations, Be stars with optically thin free-free emission and intermediate mass stars likely harboring
disks; 2) debris disk excesses are more common at 24$\mu$m than at shorter wavelengths; and 3) there
are several extreme disks similar to the nearby young debris disks around $\beta$ Pic, HR 4796A, and
49 Cet.  Finally, in \S 4 we place h and $\chi$ Per in the context of results for other open 
clusters/associations with optically-thin debris disk candidates.  
The flux from debris disks rises from $\sim$ 5 Myr (when they first emerge), 
peaks at $\sim$ 10-15 Myr, and then falls as t$^{-1}$ as described by R05.  We conclude 
with a summary of our findings and discuss future observations that may place even stronger 
constraints on debris disk evolution by accounting for wide range of IR excesses at 
10-15 Myr.
\section{Observations}
\subsection{MIPS and Ground-Based Spectroscopic Data}
We acquired MIPS 24 $\mu m$ data using 80-second exposures in 
scan mode, covering two 0.3 square-degree regions centered on the two clusters. 
The frames were processed using the MIPS Data Analysis Tool
(Gordon et al. 2005). PSF fitting in the IRAF/DAOPHOT package was used to
obtain photometry using a 7.3 Jy zero-point for the 24 $\mu$m magnitude
scale.  The typical errors for the MIPS sources are 0.2 mag ($\sim$ 5$\sigma$) at 
a 24$\mu m$ magnitude of [24] $\sim$ 10.5-11.  The number counts for the 
MIPS data peak at [24] $\sim$ 
10.5 and decline to zero by [24]$\sim$ 11.5 (Figure 1a).  
We detect 2,493 potential h and $\chi$ Per sources.  

We combined the MIPS photometry with the 2MASS/IRAC catalogue of h and $\chi$ 
Persei from C07a.  
To minimize potential contamination of stellar sources by background 
PAH-emission galaxies and AGN, we used a small 1.25" matching radius (about half of a MIPS pixel; r$_{M}$)
to merge the 2MASS/IRAC and the MIPS catalogues.  Although the MIPS beam is 6
arcsec in diameter, the instrument delivers positions good to one arcsec even for faint sources in crowded
fields (Bai et al. 2007).  This procedure yielded 616 sources (N$_{MIPS}$) with high-quality 1--24 $\mu m$ photometry.
Table 1 shows the 2MASS/IRAC + MIPS catalogue.  Optical UBV photometry from Slesnick et al. (2002) is 
included where available.

To [24]=10.5, the probability of chance alignments between 
distant PAH-emission galaxies/AGN and our sources is low.  Using the galaxy number counts from 
Papovich et al. (2004), N$_{G}$ $\sim$ 3.5$\times$10$^{6}$/sr, 
we derive a probability of $\sim$ 24.8\% that one of our 616 sources is contaminated 
($\pi$r$_{M}$$^{2}$$\times$N$_{G}$$\times$N$_{MIPS}$/(3282.8$\times$3600$^{2}$)).  The likelihood 
that many of our sources are contaminated is then much smaller.

To estimate the completeness of the MIPS sample, we compare the 
fraction of J band sources detected with MIPS within either cluster. 
Figure 1b shows that $\gtrsim$ 90\% of the 2MASS sources brighter than J=10.5 
are also detected in MIPS.  The completeness falls to $\sim$ 50\% by J=11 
and to $\sim$ 10\% by J=12.  The dip at J$\sim$8-9 occurs because many sources in this range 
are near the cluster centers, where the high density of even brighter sources (J $\sim$ 6-8) 
masks the presence of fainter objects.

To provide additional constraints on the 24$\mu m$ excess sources, 
we also obtained Hectospec (Fabricant et al. 2005) and FAST (Fabricant et al. 1998) spectra of selected 
 MIPS sources on the 6.5m MMT and 1.5m Tillinghast telescopes at F. L. Whipple Observatory
during September-November 2006.  Spectra for bright sources (J $\le$ 13) were also cross referenced with 
the FAST archive.  The FAST spectroscopy, described in detail by 
Bragg \& Kenyon (2002), typically had $\sim$ 10 minute integrations using
a 300 g mm$^{-1}$ grating blazed at 4750 $\AA$ and
a 3$\arcsec$ slit.  These spectra cover 3700--7500 \AA\ at 6 \AA\ resolution.
The typical signal-to-noise ratios were $\gtrsim$ 25-30 at 4000 \AA.
For each Hectospec source, we took three, 10-minute exposures
using the 270/mm grating.  This configuration yields spectra at
4000-9000$\dot{A}$ with 3$\dot{A}$ resolution.  The data were
processed using standard FAST and Hectospec reduction pipelines (e.g. Fabricant et al. 2005).

We acquired additional spectra of h and $\chi$ Per sources with 
the Hydra multifiber spectrograph (Barden et al. 1993) on the WIYN 3.5 m telescope at the Kitt Peak National 
Observatory.  Hydra spectra were obtained during two observing runs in November 2000 and October 2001 
and include stars brighter than V=17.0.  We used the 400 g mm$^{-1}$ setting blazed at 42 degrees, with 
a resolution of 7 \AA\ and a coverage of 3600-6700 \AA.  
The standard IRAF task \textit{dohydra} was 
used to reduce the spectra.  These spectra had high signal-to-noise with 
$\gtrsim$ 1000 counts over most of the wavelength coverage.  

\subsection{Spatial Distribution of MIPS sources}
To investigate the spatial distribution of the MIPS sources and the likelihood that they are cluster members, 
we compare the projected sky surface 
densities derived from MIPS and 2MASS.  C07a showed that $\sim$ 47\% of stars within 15' of the cluster 
centers are cluster members.  Between 15' and 25', $\sim$ 40\% of the 2MASS sources are in a halo 
population with roughly the same age as bona fide cluster stars.  Because the MIPS coverage is complete only 
out to $\sim$ 15' away from each cluster center, we cannot identify MIPS sources with this halo 
population.  We compare the spatial distribution of MIPS sources to those in 2MASS from C07a by calculating the 
number density of sources in 5'-wide half-annuli facing away from the midpoint of the two clusters.

Through 15' away from either cluster center, 
the number counts of sources detected with both MIPS and 2MASS fall off about as steeply or slightly more steeply than the 
counts for 2MASS alone from C07a (Figure \ref{dens}).  Near the center of the clusters the  
density of MIPS sources is $\sim$ 0.4/sq. arc-minute, or about an order of magnitude lower than from 2MASS.  
For h Persei and $\chi$ Persei, respectively, this density falls off by 4 \% and 25 \% from 0-5' to 5'-10' away
from the cluster centers and 22\%-41\% from 0-5' to '10-15' away from the centers.  The low counts 
through 5' and more shallow drop in number density for $\chi$ Persei is due to crowding in the inner 
$\sim$ 1-2' of the $\chi$ Persei core; the slope of the MIPS number density in $\chi$ Persei shown in 
Figure \ref{dens} is most likely a lower limit.  In contrast, the 
number counts for the 2MASS data from C07a fall off by 10\% (20\%) and 30\% (32\%) for h ($\chi$) Persei 
over the same 5' intervals (the values in Figure \ref{dens} are slightly different due to the larger annuli used here).
   The MIPS source counts appear to be about as centrally concentrated as the 
2MASS counts.

\subsection{General nature of the 24 $\mu m$ sources}

Figure \ref{k24dist} shows the histogram of K$_{s}$-[24] colors for the MIPS detections with 
2MASS counterparts.  The histogram has a main peak at K$_{s}$-[24]$\sim$0-1 and two 
groups with K$_{s}$-[24] $\sim$1-2 and K$_{s}$-[24]$\sim$2-6.
The sources with very red K$_{s}$-[24] colors ($\ge$ 2) are 
in two main groups (Figure \ref{kexc}).  A bright group of very red sources
has K$_{s}$$\sim$9-11; a fainter population of red sources 
stretches from K$_{s}$$\sim$13.5-15.  
A population of 13 Myr-old stars in h \& $\chi$ Per with 
spectral types later than B9 (M$\le$3.0 $M_{\odot}$) 
should have J, $K_{s}$ magnitudes $\gtrsim$ 13.3 (Siess et al. 2000).  
Thus, some of these fainter sources with red K$_{s}$-[24] 
colors are possibly pre-main sequence stars.

Some of the sources are very faint in the near infrared.  To examine the nature of the 
MIPS sources without J counterparts, we first compared 
the MIPS mosaic and the 2MASS J mosaic by eye.  Many sources appear scattered throughout the MIPS mosaic 
but do not appear in 2MASS, even at very high contrast.  These sources are likely 
cluster stars with J $\gtrsim$ 16-17 and very red J-[24] colors or background galaxies with 
negligible near-IR emission.  Using the number density 
of galaxies in the MIPS 24 $\mu m$ filter from Papovich et al. (2004), we expect 
$\gtrsim$ 600-700 galaxies in the 0.6 square-degree coverage area brighter than 
[24] $\sim$ 10.5.   Thus, many of the sources without 2MASS counterparts are likely not 
h \& $\chi$ Per members.  From the C07a survey, there are $\sim$ 4700
stars with J $\sim$ 10-15.5 within either cluster or the surrounding halo population of comparable age.  
For a reasonable IMF (e.g. Miller \& Scalo 1979), 
we expect $\sim$ 2800 ($\sim$ 8900) cluster/halo stars with J$\sim$16-17 
(17-18).  If $\sim$ 10\%-20\% of these stars have large 24$\mu m$ excesses, as predicted 
from an extrapolation of the C07a results to fainter J magnitudes, we expect 1170-2340 cluster/halo
stars with MIPS detections and no 2MASS counterparts.  Together with the 600-700 background 
galaxies, this population yields the observed number of MIPS detections without 2MASS
 counterparts. 

\section{A [24] IR excess population in h and $\chi$ Persei}
\subsection{Groups in the J, J-H Color-Magnitude Diagram}
To identify the nature of the 24$\mu m$ emission in sources with red K$_{s}$-[24] colors in Figure 2, 
we refer to previous MIPS observations of very 'red' sources and consider possible contaminants.
MIPS observations of the Pleiades (Gorlova et al. 2006) guide our analysis of $K_{s}$-[24] colors 
for IR excess disk/envelope sources.  While the stellar density in h and $\chi$ Per is larger than
 in the Pleiades, other possible contaminants are less important.
The level of galactic cirrus for h and $\chi$ Persei is much lower than for the Pleiades: 17-27 MJy/sr versus 
36-63 MJy/sr.  Gorlova et al. found that disk-bearing candidate sources have dereddened 
$K_{s}$-[24] colors $\gtrsim$ 0.25.  Because h \& $\chi$ Persei has a low, uniform extinction of $A_{V}\sim 1.62$, 
E(B-V)$\sim$0.52 (Bragg \& Kenyon 2005), we convert the dereddened $K_{s}$-[24] excess 
criterion into a reddened $K_{s}$-[24] criterion
using the reddening laws from Indebetouw et al. (2005) and Mathis (1990).  For
 $A_{V}$$\sim$1.62, 24$\mu m$ excess sources should have $K_{s}$-[24]$\gtrsim$0.45.  
Because the MIPS data have $\sigma$ $\lesssim$ 0.2, we round this limit up to $K_{s}$-[24] $\gtrsim$ 0.65.

Figure \ref{jjh} shows the distribution of sources with and without $K_{s}$-[24] excess in J/J-H color-magnitude space.   
The IR excess population is clustered into two main groups.  
Asterisks (diamonds) denote sources brighter (fainter) than J=13.  
Larger asterisks/diamonds correspond to sources with 
$K_{s}$-[24] $\ge$ 2, while smaller asterisks identify sources with $K_{s}$-[24]=0.65-2 and J $\le$ 13.
The excess sources with J$\le$13 typically have J-H colors $\sim$ 0.2 mag redder than 
a typical stellar photosphere.  Many sources with weak excess lie well 
off the 14 Myr isochrone and may be consistent with foreground M stars or supergiants.  
About 17 out of 21 stars with 24$\mu m$ excess fainter than J=13 fall 
 along the 14 Myr isochrone with J$\approx$ 14-15.  
At 14 Myr and a distance of 2.4 kpc, this J magnitude range corresponds to stars with masses $\sim$ 2.2-1.4 $M_{\odot}$ 
(B9/A0-G2) stars (Siess et al. 2000).  We inspected each faint excess source on the MIPS mosaic for 
extended emission (indicative of galaxies) or 'excess' due to source confusion/crowding.  We found no evidence 
issues such as as large extended emission or source confusion that could compromise the photometry of any of the faint excess sources.
\subsection{Two Populations of IR-Excess Sources: Be stars with Circumstellar Envelopes and Faint Pre-Main Sequence
 Stars with Disks}
There are three main possibilities for the source of 24$\mu m$ excess emission around h and $\chi$ Per stars.  
Red giants or supergiants not associated with the clusters or the halo produce IR excesses in massive stellar winds, 
and should have J-H $\gtrsim$ 0.5.  Two such stars have large 24 $\mu m$ excesses 
and are not considered further.  Be stars in the clusters/halo population have IR excesses from optically-thin 
free-free emission and should have J $\lesssim$ 13-13.5 and J-H $\lesssim$ 0.4.  Many potential Be stars have $K_{s}$ -[24] $\sim$ 2 
(Figure \ref{jjh}; large asterisks) and clearly are an important part of the cluster population.  Aside from Be stars, 
circumstellar disks around lower-mass cluster/halo stars can produce excess emission.  Figure 5 shows a significant 
population of fainter stars (all with J $\gtrsim$ 13.5) on the 14 Myr isochrone with large K$_{s}$ -[24] excesses.
To identify the nature of the 24$\mu m$ excess sources, we analyze the near-IR colors and selected spectra 
of the excess population.  We begin with Be star candidates and then discuss the fainter population.

Be stars are massive and are evolving off the main sequence (McSwain and Gies 2005).  The IR-excess emission from Be 
stars arises from an optically-thin, flattened, circumstellar shell of ionized gas ejected from the star (Woolf, Stein, and 
Strittmatter 1970; Dachs et al. 1988).  We find 57 candidates -- J $\lesssim$ 13.5 and J-H $\le$ 0.4 --
 with 24 $\mu m$ excess (K$_{s}$-[24] $\gtrsim$ 0.65).  Twenty of these stars have been previously identified 
as Be stars by Bragg and Kenyon (2002), all with Oosterhoff (1937) numbers, and have spectral types  
from Strom and Wolff (2005) and Bragg and Kenyon (2002).  Table 2 lists the  properties of these 57 candidates.

We can estimate the ratio of Be stars to B stars over a narrow range of spectral types (earlier than B4).
In our MIPS survey, there are $\sim$ 175 stars that are likely B type stars (J=8-13.5; J-H $\le$ 0.2) without excess. 
All of these stars are probably earlier than B4 based on their 2MASS J band photometry (J $\lesssim$ 11.75).
There are 57 Be star candidates with 24$\mu m$ excess (K$_{s}$-[24]$\gtrsim$0.65): 51 of these stars probably 
have spectral types earlier than B4 based on their J-band photometry.  The ratio of Be star  
candidates to main sequence B-type stars earlier than B4 in the MIPS survey is then $\sim$ 0.29.  
This estimate is larger than the ratio derived from optical and near-IR data ($\sim$ 0.14, Bragg \& Kenyon 2002).
We explored this difference as follows.

First, we analyze the Be star candidate population in high-density regions close to the cluster centers 
where spectroscopic data on bright stars in h and $\chi$ Per is complete. 
Our candidates were cross correlated with spectroscopically identified Be stars from Bragg and Kenyon within 
5' of the cluster centers..  Ten known Be stars and two candidate  
stars from the MIPS survey are in this region.  
These two candidate stars with 24 $\mu m$ excesses are not Be stars.
One is a 5th magnitude B3 supergiant cluster member  
identified by Slesnick et al. (2002) and the other is a G1 star, lying well off the isochrone with J=11.67 and J-H=0.39.    

Because our spectroscopic sample of Be stars is spatially limited, we turn to near-IR colors from 2MASS to 
investigate the nature of the bright MIPS excess sources.
Dougherty et al. (1991, 1994) showed that Be stars follow a distinct locus in JHK$_{s}$ colors.   
This locus is characteristic of free-free emission from optically-thin ionized gas and is well separated 
from main sequence colors and the near-IR colors produced by warm dust.
Thus, the J-H/H-K$_{s}$ color-color diagram (Figure 6) provides a clear way to distinguish Be stars from 
lower mass stars with circumstellar dust emission.
From Figure 6, it is clear that the bright sources with K$_{s}$-[24]$\ge$2
 follow a locus (dotted line) in J-H/H-$K_{s}$ from 
(0,0) to (0.3,0.4), a range consistent with known Be star colors 
(Dougherty et al. 1991, 1994).  The bright sources with weaker excess (small asterisks) also appear to 
 lie along the Be star locus or are clumped close to the red giant locus at J-H $\sim$ 0.6-0.8, 
H-$K_{s}$$\sim$ 0.2-0.3.  The observed distribution of IR colors 
suggests $\approx$ 15 Be stars and $\approx$ 25 giants/supergiants.  
If this ratio is confirmed by optical spectroscopy, then the fraction of Be stars among 
all B-type stars is similar to the 14\% derived by Bragg \& Kenyon (2002).  

Finally, we search the FAST archive at the Telescope Data Center at the Smithsonian Astrophysical 
Observatory and the Slesnick et al. catalogue for additional spectra of the 35 Be star candidates in lower-density regions.
The FAST archive contains additional data for four candidates; we find one additional source from Slesnick et al.  
The Slesnick et al. source is a confirmed Be star (B1Ie).  The FAST sources contain an A2, F7, G2, and B4 star.  The first 
three of these are bright and likely either foreground or giants associated with the halo population of h and $\chi$ Per.
Thus, the spectra support our conclusion from the color-color diagram (Figure 6) that 
many of the candidate Be stars are not true identifications.  If none of the remaining candidate stars are true 
Be stars, then the ratio of Be stars to B stars is $\sim$ 0.12, close to the Bragg and Kenyon value. 

Interestingly, the B4 star identified by Bragg \& Kenyon (2002), has 24 $\mu m$ excess 
\textit{and} has a J magnitude and J-H colors marginally consistent with an early B star in h and $\chi$ Per.
The $K_{s}$-[24] excess for this source is $\sim$ 1.05, though unlike Be stars it lacks clear IR excess at JHK$_{s}$ and 
in the IRAC bands.  We show its spectrum compared with that of a known Be star (Oosterhoff number 517) in Figure 7.  
This star is the earliest, highest mass star+circumstellar disk source known so far in h \& $\chi$ Per.

Based on their near-IR colors and optical spectra, the faint excess sources in our survey are clearly 
distinguishable from Be stars.  The near-IR colors of the faint excess sources are
 evenly distributed between J-H = 0.1-0.6 and H-$K_{s}$=0-0.2 
(Figure 6, diamonds).  The lack of very red H-K$_{s}$ colors for a typical faint excess 
source is consistent with a lack of warm (T$\sim$ 1000 K) circumstellar envelope emission.
Nearly all (17/21) faint sources are photometrically consistent with h \& $\chi$ Per membership, though
high-quality spectroscopic data are currently limited to 8 sources (2 from Hydra, 6 from Hectospec).  
Figure 8 shows the spectra.
Seven of the eight faint excess sources have spectra consistent 
with h and $\chi$ Per membership.  The one non-member source in Figure 8
is an F5 star with J=15.9, beyond the 2MASS 
completeness limit and below the isochrone by $\sim$ 1.0 magnitude, and thus 
is one of the four faint sources that is also photometrically inconsistent with cluster 
membership.  
The spectral types for the 7 sources consistent with cluster membership 
range from A2 to F9.  None show strong H$_{\alpha}$ emission which is a 
signature of accretion (e.g. White \& Basri 2003) and thus a reservoir of 
circumstellar gas.   These stars are therefore 
very similar to the nearby young (8-12 Myr old) 
debris disks $\beta$ Pic, HR 4796A, and 49 Cet in that they have comparable 
spectral types, have 24 $\mu m$ excess, and lack any signatures of gas accretion.
The 13 sources without known spectral types are either unobserved (12) or had too low  
signal-to-noise to derive spectral types (1, this source is well off the isochrone).  

Many of the fainter 24$\mu m$ sources have high-quality IRAC photometry.
Of the 17 MIPS excess sources fainter than J=13.5 
on the isochrone, 14 (12, 11) also have IRAC 
measurements at [4.5] ([5.8], [8]).  We summarize the observed properties of 
the faint MIPS excess sources in Table 3.  In the following two sections, we focus 
on these sources, comparing the MIPS photometry to IRAC/2MASS photometry from C07a and 
modeling the sources' emission from 2.2 $\mu m$ through 24$\mu m$.
\subsection{Nature of the Disk Population in Faint Pre-Main Sequence Stars}
\subsubsection{Mid-IR colors and the Wavelength-Dependent Frequency of Disks}
To constrain the nature of the 17 faint MIPS excess sources that are (photometrically) 
consistent with cluster membership, we compare the K$_{s}$/$K_{s}$-[24] CMD with CMDs using three IRAC colors, 
$K_{s}$-[4.5], K$_{s}$-[5.8], and $K_{s}$-[8] in Figure \ref{cmdirac} (diamonds).  For reference, 
we also show the colors for bright MIPS sources without 24$\mu m$ excess (squares).
Following C07a, we identify sources with $K_{s}$-[IRAC] colors $\ge$ 0.4 as 
IR excess sources; sources with $K_{s}$-[24]$\ge$0.65 are 24$\mu m$ excess 
sources.  A vertical line in Figure \ref{cmdirac} shows the division between excess and non excess sources.

The frequency of IR excess varies with wavelength.
Only 1/14 faint 24$\mu m$ excess sources also have excess at [4.5].  The fraction of sources with 
[5.8] excess is 3/12.  The 8$\mu m$ excess population has a larger fraction of 
excess sources, 5/11.  While some of the 'photospheric' sources, $K_{s}$-[4.5, 5.8, 8] $\le$ 0.4, 
may have weak excesses, many sources have $K_{s}$-[IRAC] $\le$ 0.2 (observed) and $\lesssim$ 0.1 (dereddened).  
These sources are unlikely to have any dust emission at [4.5], [5.8], or [8].  
While the small sample of 24$\mu m$ excess sources precludes a strong statistical significance for any trend 
of IR excess emission, the wavelength-dependent frequency of excess emission is consistent with 
 results from larger surveys (e.g. C07a; Su et al. 2006).
\subsubsection{Temperature and Location of Circumstellar Dust}
Analyzing the strength of IR excess emission at multiple bands 
 places constraints on the temperature and location of the dust.
Just over half of the faint 24$\mu m$ excess sources have no
excess emission in the IRAC bands, so these sources 
lack circumstellar material with temperatures $\gtrsim$ 400K.  
Because a blackbody that peaks at 24$\mu m$ has T $\sim$ 120 - 125 K, the dust 
temperature in most of the faint 24$\mu m$ excess sources is probably $\lesssim$ 100-200 K. 

We can put more quantitative constraints on the dust temperature with
  a flux ratio diagram.  Flux ratio diagrams have been an important
 tool in analyzing accretion disks in unresolved cataclysmic 
variable systems (e.g. Berriman et al. 1985; Mauche et al. 1997).  In this 
method, the ratio of fluxes (in this case, $\lambda$F$_{\lambda}$) at different wavelengths 
 such as $\lambda_{4.5}$F$_{4.5}$/$\lambda_{8}$F$_{8}$ and $\lambda_{24}$F$_{24}$/$\lambda_{8}$F$_{8}$ is computed.   The ratios 
for blackbody emission follow a curve in flux ratio space.  Because disk-bearing sources should 
be, to first order, the sum of two blackbodies (a hot stellar component and a cooler circumstellar 
component), their positions in flux ratio space should lie on a line between the circumstellar dust temperature 
and the stellar temperature.  

Figure \ref{fr} shows the flux ratio diagram for our sample, and  Table 4 lists the derived 
disk temperatures (labeled as T$_{D}$ FR).  
We restrict our sample to 10 sources with 5$\sigma$ detections from 4.5 through 
24 $\mu m$\footnote{The first source in Table 3, with J=13.84, has a
 [8] flux that has a negative K$_{s}$-[8] color and thus is unphysically faint.  
An unphysically large ratio of the [4.5] to [8] flux cannot be interpreted with 
a flux-ratio diagram.}.
Five of these sources have [8] excess; one has [4.5] excess.  
For 13-14 Myr-old sources, the range of spectral types with J=14-15.5 is  
$\sim$ A0 to G8 (Siess et al. 2000).
The flux ratios for blackbody emission from 10 K to 10000 K follow the solid line with 
the temperatures characteristic of disks ($\sim$ 10-1000 K) on the vertical part of the line
and those for stellar photospheres on the horizontal part.
Loci showing the locations for a stellar photosphere+disk of a given temperature are shown 
 ranging from T$_{disk}$= 300 K to 100 K assuming a stellar temperature of 
$T_{e,\star}$$\sim$ 7250 K (about F0 spectral type).  The sources without (with) IRAC excess emission, 
K$_{s}$-[IRAC] $\lesssim$ 0.4, are shown as diamonds (thick diamonds).  The further away
from the origin point of the loci a given source is, the more the disk contributes to the total flux. The 
derived dust temperatures are only weakly sensitive to $T_{e,\star}$ as
flux ratios for 5250-10000 K ($\sim$ G9-B8) blackbodies occupy roughly 
the same place at $\lambda_{4.5}$F$_{4.5}$/$\lambda_{8}$F$_{8}$ ($\sim$ 5$\pm$0.15) 
and $\lambda_{24}$F$_{24}$/$\lambda_{8}$F$_{8}$ ($\sim$ 0).  
The line for the ice sublimation temperature is shown in bold.
The source (diamond at $\sim$ 2, 0.5) with a disk 
component of $\approx$ 300 K has a strong [8] excess and was
  previously identified as having $\sim$ 300-350 K dust (Source 5 in C07b) using a 
single blackbody $\chi$$^{2}$ fit to the disk SED.  
Four other sources, also with 8 $\mu m$ emission, have dust temperatures between 230 K and 250 K.  
All the sources with 8$\mu m$ excess then have dust temperatures $\ge$ 230 K.
While these sources may have cooler dust components, some of the dust emission must 
come from warmer disk regions closer to their parent stars.

The dust temperatures of sources without 8$\mu m$ excess, characteristic of a slight majority 
 in our sample, are significantly lower.
Three sources have slightly cooler temperatures of 
$\sim$ 170-185 K, comparable to the water ice sublimation temperature (Hayashi 1981).
The remaining sources have much cooler dust temperatures ($\sim$ 100-150 K). 
This diagram demonstrates that many sources 
must have cold dust with temperatures of T$_{dust}$ $\lesssim$ 200K.

For sources with photospheric IRAC emission, using a single-temperature 
blackbody - calculated by matching the 24 $\mu m$ excess while not producing significant excess in the IRAC bands - should 
match the observed disk emission well.  However, many sources also have IRAC excess, and modeling the disk emission as coming 
from two sources (e.g. warm \textit{and} cold dust) may yield a significantly better fit (e.g. 
Augereau et al. 1999).  As an alternate way to constrain the disk temperature(s) and 
estimate the disk luminosity and location of the dust, we now consider blackbody fits to the dereddened SEDs.  
For sources with IRAC excess we add sources of hot and cold dust emission with temperatures of 50-250 K and 250-700 K, respectively, 
to the stellar photosphere.  Sources without IRAC excess are modeled by a stellar photosphere + single-temperature disk.
For the stellar blackbody, we use the conversion from spectral type to effective temperature from Kenyon and Hartmann (1995).  
If the star has no spectral type, we use the dereddened J band flux as a proxy for spectral type as in C07a, assuming 
A$_{J}$ $\sim$ 0.45 and using the Kenyon and Hartmann (1995) conversion table, and add a question mark after the spectral type 
in Tables 3 and 4.  We use the stellar luminosity, L$_{\star}$, for 13-14 Myr old stars of a given spectral type from Siess et al. (2000). 
A $\chi$$^{2}$ fit to the 3.6-24 $\mu m$ fluxes is performed to find the best-fit one or 
two dust blackbody + stellar blackbody model following Augereau et al. (1999).  
Following Habing et al. (2001), we derive the disk luminosity from blackbody fits.    
The integrated fluxes for each dust population of a given temperature are added and then divided by 
the stellar flux to obtain the fractional disk luminosity, L$_{D}$/L$_{\star}$.  Finally, we estimate the location of the dust 
populations from simple radiative equilibrium:
\begin{equation}
R(AU) \approx (T_{disk}/{280})^{-2}({L_{\star}}/{L_{\odot}})^{0.5}.
\end {equation}

The sources without IRAC excess have nearly identical disk temperatures to those derived from the 
flux-ratio diagram (Table 3), ranging from $\sim$ 90 K to 185 K, and are similar to equilibrium temperatures 
just beyond the terrestrial zone into the gas giant regions of the solar system.  These sources have $\chi$$^{2}$ 
values slightly less than or comparable to the number of observations ($\sim$ 1-6).  The fractional disk luminosities range from 
$\sim$ 5.5$\times$10$^{-4}$ to 3.5$\times$10$^{-3}$, which is similar to dust luminosities for young stars surrounded by optically-thin 
debris disks (e.g. Meyer et al. 2007).  Dust in these systems is probably confined to disk regions 
of $\sim$ 8-40 AU.

Sources with both IRAC and MIPS excess emission have disk temperatures substantially different from 
those inferred from the flux-ratio diagram  and show evidence of terrestrial zone dust emission and colder dust.  
The two dust population fits for the IRAC+MIPS excess sources show evidence 
for a wide range of dust temperatures with warm terrestrial dust emission and cold dust 
emission similar to that from sources without IRAC excess.  For instance, 
the SED of the A6 star with IRAC and MIPS excess is best fit ($\chi$$^{2}$ $\sim$ 7.6) by a hot dust component of 375 K coming from 
1.8 AU and a cold component of 85 K at $\sim$ 37 AU.  The faint F9 star, identified previously as 'Source 5', 
is extremely well fit ($\chi$$^{2}$ $\sim$ 0.5) by dust populations of 240 and 330 K at 1.8 and 3.4 AU, respectively.  
Because these sources have both warm and cold dust, it is not surprising that 
their fractional disk luminosities are typically higher.  The fractional luminosity of Source 5 ($\sim$ 6$\times$10$^{-3}$) is 
comparable to the most massive debris disks (e.g. HR 4796A), and in general the luminosity of the disk population is 
consistent with values for massive debris disks.  The most luminous disk source 
(L$_{D}$/L$_{\star}$ $\sim$ 1.5$\times$10$^{-2}$) is the lone exception and 
has a luminosity halfway in between values expected for luminous debris disks ($\sim$ several$\times$10$^{-3}$) and long-lived T Tauri 
disks with inner holes (e.g. TW Hya; Low et al. 2005).  We analyze this system further in \S 3.3.3.

In summary, the faint MIPS excess sources have dust with a range of temperatures and luminosities.
Sources without IRAC excess are well fit by single-temperature blackbodies and have cold dust components with temperatures 
$\sim$ 90-185 K.  Sources with IRAC excess are better fit by two dust components, a hotter, terrestrial zone component 
and a cooler component.  The disks in h and $\chi$ Persei then show evidence of having inner regions of varying 
sizes cleared of dust.  All but one source has a fractional disk luminosity $\lesssim$ 10$^{-2}$, consistent with optically-thin 
debris disks.  In the next section, we investigate the evolutionary state of the faint MIPS excess population further 
by comparing their properties to other predicted properties for massive debris disks and T Tauri disks.
\subsubsection{Evolutionary State of the MIPS Disk Candidates: A Population of Luminous $\sim$ 13-14 Myr-old Debris Disks}
We now consider the evolutionary state of the dust in the 24$\mu m$ excess sources.  
Although the relative luminosities (L$_{d}$/L$_{\star}$ 
$\sim$ 10$^{-3}$) and lack of accretion signatures suggest these h and $\chi$ Per sources are debris disks, some T Tauri stars (e.g. 
'transition' T Tauri stars; Kenyon \& Hartmann 1995) may also have inner regions cleared of gas and dust.  
Thus, it is important to compare their disk properties to models of debris disks and T Tauri disks. 

We first examine the nature of the h and $\chi$ Per disk population
as a whole. Because our smallest disk luminosities,
$\sim 5 \times 10^{-4}$, are larger than more than half of
known $\gtrsim$ 10 Myr-old disks (e.g. Meyer et al. 2007), our MIPS
sample probably misses lower luminosity sources with
$L_d/L_{\star} \lesssim 10^{-4}$. Similarly, our lack of
70 $\mu$m detections limits our ability to detect and to
evaluate disk emission from cooler dust -- such as is observed in
$\beta$ Pic and HR 4796A -- with SEDs that peak at 40--100 $\mu$m.
For example, the nearby, luminous disk around 49 Cet (spectral type A1V, 8 Myr old;
\citealt{Wa07}) has a 24$\mu$m excess of $\sim$ 2.5 magnitudes.
We detect only one faint (J $\ge$ 13) MIPS excess sources with K$_{s}$-[24] $\le$ 2.5.  Therefore, it is possible that
there are additional very luminous young disks in h and $\chi$ Per just below our detection limit. 

These limits and the rich nature of the Double Cluster allow
us to estimate the prevalence of massive, luminous disks.
The total number of A0 to early F stars (F2) in h and $\chi$ Per is $\approx$ 1000 (Currie et al. 2007 
in prep.; cf. Currie et al. 2007a).  Assuming that the disk fraction is 
$\sim$ 20\%\footnote{disk fractions quoted by Chen et al. (2005b) range from 
9\% to 46\%}, we detect 17/200, or $\approx$ 10\% of all disks with strong emission at
24$\mu$m.   Thus, this population
is extreme and yields a better understanding of the evolutionary
state of the most luminous disks in a populous star cluster.

To constrain the evolutionary state of the disks, we 
compare the near-to-mid infrared disk colors to those expected for two disk models:
a flat, optically-thick disk around a Classical T Tauri star 
($T_{disk}$$\sim$$r^{-0.75}$; Kenyon \& Hartmann 1987)
and an optically-thin disk 
model from Kenyon \& Bromley (2004a) for debris emission produced by planet formation. 
Because only one of our sources has [4.5] excess emission and less than half 
have [8] excess emission, we match the data to models of planet formation not in the 
terrestrial zone (Kenyon \& Bromley 2004a) but at 30-150 AU from a 2.0 $M_{\odot}$ 
primary star (Kenyon \& Bromley 2004b).  For a $\sim$ 2.0 $M_{\odot}$, 20 L$_{\odot}$ star, the temperature range from 30 to 150 AU 
is comparable to the outer gas/ice giant region in our solar system ($\sim$ 6.7-34 AU).
We adopt a $\Sigma \propto$ $r^{-1.5}$ profile for the initial column density 
of planetesimals and an initial disk mass of 3$\times$ a scaled Minimum Mass Solar Nebula (Hayashi 1981): 
3$\times$ 0.01 M$_{\star}/M_{\odot}$ (where M$_{\star}$ = 2 M$_{\odot}$).
Emission from planetesimal collisions is tracked for $\sim$ 10$^{8}$ yr.  Model predictions are reddened  
to values for h and $\chi$ Persei (reddening laws in the IRAC/MIPS bands are described in C07b).

Figure \ref{colcol} shows the K$_{s}$-[4.5, 5.8]/K$_{s}$-[24] color-color diagrams for bright photospheric 
sources and the faint 24$\mu m$ excess sources.
The debris disk locus is overplotted as a thin black line.  Debris from planet formation  
produces a peak excess emission at $K_{s}$-[24] $\sim$ 3.6 at $\sim$ 10$^{7}$ years; the $K_{s}$-[4.5] 
and $K_{s}$-[8] colors peak at $\sim$ 0.4 at earlier times ($\sim$ 10$^{6}$ years).  The debris disk locus tracks the colors 
for most of the sources in $K_{s}$-[4.5]/$K_{s}$-[24] space very well 
(Figure \ref{colcol}a).  While the locus underpredicts the [8] excess for about half of the 
sources (Figure \ref{colcol}b), warmer regions of a debris disk 
not modeled here may produce this excess (e.g. KB04a).  C07b showed that  planet formation in the 
terrestrial zone can produce strong [8] emission characteristic of some h and $\chi$ Per 
sources at $\sim$ 10-15 Myr.  Indeed, the source with $K_{s}$-[8]$\sim$ 1.3, 
$K_{s}$-[24]$\sim$ 4.4 is Source 5 from C07b which was one of eight modeled as having terrestrial zone 
debris disk emission.  The warm dust temperature ($\sim$ 300 K) derived for this 
source in \S 4.2 is consistent with terrestrial zone emission.

Disk models corresponding to earlier evolutionary states fare worse in matching the observed mid-IR colors.
 The optically-thick flat disk model (the triangle in both plots)
predicts $K_{s}$-[5.8] ([8]) $\sim$ 1.5 (2.9) and $K_{s}$-[24]$\sim$6, consistently 1-2 magnitudes redder 
than the data.  To match the observed [24] excess, any
optically-thick disk with an inner hole (cf. C07b) must be cleared of dust out to the distances probed 
by the MIPS bands: $\sim$ 25 AU for a 20 $L_{\odot}$ primary star.  While inner hole models may be 
constructed to fit the SEDs of sources with only 24 $\mu m$ excess, these models predict nearly zero 
IRAC color even though about half of the sample has excess at [8].  Lack of gas accretion signatures, 
low fractional disk luminosities, and SED modeling then suggest that at least many faint h and $\chi$ 
Per sources with 24 $\mu m$ excess are stars surrounded by optically-thin debris disks.
More sensitive spectroscopic observations are needed to verify the lack of gas in these 
systems.

Despite the general success of the debris disk models, at least one h and $\chi$ Per source may 
harbor a disk at an earlier evolutionary state.
This source has a K$_{s}$-[24] color of $\sim$ 6, which is $\sim$ 1 mag redder than HR 4796A, 
the strongest 24$\mu m$ excess source in R05.  This color is close to the optically-thick disk predictions,  
 is extremely difficult to produce with a debris disk model, and is more similar to the level of excesses in 
 older T Tauri stars like HD 152404 and TW Hya (Chen et al. 2005b; Low et al. 2005).  

To explore this possibility, we 
overplot the K$_{s}$-[4.5], K$_{s}$-[8], and K$_{s}$-[24] colors of TW Hya from Hartmann et al. (2005) and 
Low et al. (2005) in Figure \ref{colcol} (large cross, reddened to h and $\chi$ Per).  The mid-IR colors of our brightest source are 
similar to the colors of TW Hya.  While TW Hya's disk has an optically-thin inner region 
where the early stages of planet formation may be commencing (Eisner et al. 2006), the disk is probably optically-thick 
at 24$\mu m$ (Low et al. 2005).  TW Hya also has strong H$_{\alpha}$ emission which 
indicates accretion.   On the other hand, the fractional disk luminosity in h and $\chi$ Per sources
is much lower than that of TW Hya ($\sim$ 0.27, Low et al. 2005) and in between values for debris disks 
and transition disks.  Thus, some lines of evidence suggest that this extreme h and $\chi$ Per source 
is at an evolutionary state earlier than the debris disk phase while others are more ambiguous.
Obtaining optical spectra of this source, to search for accretion signatures, may allow us to make a better 
comparison between it and older T Tauri stars like TW Hya.

The spectral energy distributions (SEDs) of the faint MIPS excess sources show 
evidence for a range of dust temperature distributions, which may be connected to a range 
of evolutionary states (Figure \ref{sed4}).  We select four sources, three with spectra and 
one without, that are representative of the range of mid-IR colors from our sample.
The first three sources of Figure \ref{sed4}, dereddened to A$_{V}$=1.62 (E(B-V)=0.52),
 have been spectroscopically confirmed as 
F9 (source 1), F9 (source 2), and A6 (source 3) stars, respectively; the second source was mentioned in the previous 
paragraph (with K$_{s}$-[24]$\sim$ 4.4).  The SEDs 
for the bottom left source was also dereddened to A$_{V}$=1.62, and a spectral type 
of A2 was chosen based on the conversion from absolute magnitude to spectral type 
for 14 Myr-old sources (from Siess et al. 2000; Kenyon \& Hartmann (1995) color conversions).
The source with photospheric emission at $\lambda$ $<$24$\mu m$ (source 4) has 
IRAC colors representative of just over half of the faint MIPS-excess sources in 
Figure \ref{colcol}.  The debris disk model accurately predicts the SEDs of the source with photospheric 
8$\mu m$ emission and two sources with weak 8$\mu m$ excess emission.  The remaining source is not 
fit well by the disk model and shows clear evidence for a large warm dust population (see C07b). 
The evolutionary states for the sources shown in Figure \ref{sed4} and
  the 9 sources with complete IRAC and MIPS photometry are listed Table 2. 

Thus, we conclude that emission from at least half of the 24$\mu m$ excess sources around pre-main sequence stars 
in h and $\chi$ Per is best explained by debris from
planet formation at locations comparable to the gas/ice giant regions in the solar nebula.  Some of the other 
pre-main sequence stars with 24 $\mu m$ excess may also have ongoing planet formation in the
inner, terrestrial zone regions as indicated by their 8 $\mu m$ excesses.  One of our sources 
may be a T Tauri star at a slightly earlier evolutionary state than the debris disk sources 
in our sample.  

If most of the disk population is then interpreted as an early debris disk population 
(not a Class II/III transition T Tauri disk population), the wavelength-dependent frequency 
of IRAC/MIPS disk excess identified in \S3.3.1 implies a location-dependent evolution of 
debris disks, specifically a clearing of warm dust from inner disk regions.  
This behavior is consistent with standard models of planet formation (KB04a), 
which predict that dust emission from the planet formation process disappears at shorter 
wavelengths (e.g. IRAC bands) faster than at longer wavelengths (e.g. MIPS bands).
This result is expected if planet formation runs to completion in the innermost regions 
of protoplanetary disks before planets are formed in the outer disk.

\section{Evidence for a Rise and Fall of Debris Disk Emission}
To place our results in context, we now compare the excesses observed 
in h and $\chi$ Per sources with measurements of other stars with roughly similar masses.  We follow  
 R05 and consider the magnitude of the 24$\mu m$ excess, [24]$_{obs}$-[24]$_{\star}$, as a function of time.
Using a sample of early (A) type stars with ages $\gtrsim$ 5 Myr, R05 showed that stars have a wide range of 
excesses at all ages and that sources with the largest excesses define an envelope that decays slowly with 
time (Figure \ref{excvagegr}).  Although this envelope is consistent with a power-law decay, 
[24]$_{obs}$-[24]$_{\star}$ $\propto$ t$^{-1}$, the R05 sample has relatively few stars with ages 
$\sim$ 5-20 Myr where debris disk models predict large excesses.  The sources with the largest 
excesses, HR 4796A and $\beta$ Pic, fall within this age range at 8 and 12 
Myr\footnote{While $\beta$ Pictoris was given an age of 20 Myr in R05, derived from Barrado y Navasceus et al. (1999),
recent work suggests a slightly younger age of $\sim$ 12 Myr (e.g. Zuckerman et al. 2001; Ortega et al. 2002).}, respectively.

Together with our results for h and $\chi$ Per, several recent surveys in young clusters and associations 
identify debris disks with ages of $\sim$ 5-20 Myr (Chen et al. 2005b; Hernandez et al. 2006; 
Sicilia-Aguilar et al. 2006).  As in R05, these surveys show a large range of 24 $\mu m$ excesses 
at each age.  In the well-sampled Sco-Cen Association, for example, Chen et al. (2005b) identify many stars 
with photospheric emission (no excess) at 24$\mu m$ and several stars with excesses considerably 
larger than the typical excess observed in the R05 sample.  Although our h and $\chi$ Per data 
do not provide any measure of the number of stars with photospheric emission at 24 $\mu m$, 
the survey yields a good sample of stars with excesses much larger than the typical R05 source.  

We now combine our results with those from R05 and from more recent surveys of debris disks in young clusters.
Specifically, we add data from Tr 37 (4 Myr) and NGC 7160 (11.8 Myr) (in Cepheus; Sicilia-Aguilar et al. 2006), 
 Orion OB1a (10 Myr) and Orion OB1b (5 Myr; both from Hernandez et al. 2006), and Sco-Cen ($\sim$ 5, 16, and 17 Myr 
for Upper Sco, Lower Centaurus Crux, and Upper Centaurus Lupus, respectively; Chen et al. 
2005b).  For h and $\chi$ Per and Cepheus sources, we include only the IR-excess sources.  The sensitivity of the
Sco-Cen observations allows more precise determinations of the photospheric flux levels farther down the initial mass function of 
the cluster, so we include data for all sources earlier than G0 with or without excess in this cluster.  
For sources with no published estimate of the photospheric flux, we assume that K$_{s}$-[24]$_{\star}$ $\sim$ 0 (dereddened), 
which is valid for our sample of A and F stars.

\subsection{Observed Mid-IR Emission vs. Age}
When data from h and $\chi$ Persei and other young clusters are added to R05,
the evolution of 24 $\mu m$ excess with age shows an important new trend.  \textit{From $\sim$ 5-10 Myr, 
there is a clear rise in the magnitude of excess followed by a peak at $\sim$ 10-15 Myr, and a slow t$^{-1}$ 
decay after $\sim$ 15-20 Myr} (Figure \ref{excvage}).  All sources with very large 
($\gtrsim$ 3 mag) excesses have ages between 8 and 16 Myr.  The 24 $\mu m$ excess 
emission peaks at $\sim$ 12-16 Myr as indicated by strong excess sources in h and 
$\chi$ Persei (diamonds), NGC 7160 (squares), and Sco-Cen (asterisks).  
Data from 5 Myr-old Orion OB1b and Upper Sco to 10 Myr-old Orion OB1a to 12-17 Myr-old NGC 7160, h and $\chi$ Per, 
and the two older Sco-Cen subgroups shows a 
sequential rise in the median 24 $\mu m$ excess  \footnote{The debris disk 
candidates in Tr 37 have larger excesses than those in Orion OB1a.  However, the strong excess may be explained by 
differences in stellar properties: 2/3 of the debris disk systems in Tr 37 are B3/B5 and B7 stars, which are 
far more massive than 10 Myr-old A/F stars ($\sim$ 3.5-6 M$_{\odot}$ vs. 1.5-2.5 M$_{\odot}$; cf. Siess et al. 2000).  If typical disk 
masses scale with the stellar mass, then these much more massive stars should have more massive, more strongly emitting disks.  
The disk mass-dependent amplitude of excess is discussed in \S 4.3.}. A peak in the 24 $\mu m$ excess emission at $\sim$ 10 Myr is 
 also visible in a plot from Hernandez et al. (2006), albeit at a lower statistical significance.  
The addition of several $\lesssim$ 20 Myr-old clusters 
more strongly constrains the time when debris emission peaks and maps out its evolution from 5-20 Myr 
in more detail.

Removing possible TW Hya-like sources from this diagram does not modify the trends.
The 'TW Hya-like' source in h and $\chi$ Per ([24] excess $\sim$ 5.5) and the strongest excess source 
in Sco-Cen ([24] excess $\sim$ 5.75) have the largest 24$\mu m$ excesses and may be at an evolutionary state 
prior to the debris disk phase.  However, many sources in 
the 10-15 Myr age range have $\sim$ 2-3.5 magnitude excesses, including HR 4796A and many 
h and $\chi$ Per sources, which have disk luminosities and mid-IR colors inconsistent with an optically-thick disk.
The second most luminous source in Sco-Cen, HD 113766A (F3 spectral type) with a 24 $\mu m$ excess of $\sim$ 4.7 magnitudes,
 has a fractional disk luminosity characteristic of a massive debris disk (Chen et al. 2005b).
Sources with [24]$_{observed}$-[24]$_{\star}$ $\ge$ 2-3 are more common at $\sim$ 10-15 Myr than at much younger
($\sim$ 5 Myr) or older ($\gtrsim$ 20 Myr) ages.

\subsection{Statistical Verification of a Peak in 24 $\mu m$ Emission at 10-15 Myr}
The peak at 10-15 Myr is statistically robust. To test it, we adopted the underlying approach that Wyatt et
al. (2007a) demonstrate gives a good first-order description of debris disk behavior: debris disks all evolve
in a similar fashion, with the variations among them arising primarily from differences in initial mass. This
result has two important implications for us: 1.) it validates deducing evolution with time from the upper
envelope of the infrared excesses, since similar high-mass disks of different ages define this envelope; and
2.) it allows us to estimate the distribution of excesses at any time by scaling the excesses at another time
according to $t^{-1}$ (by one over the ratio of the source ages), the general time dependence of disk decay (R05). We use the second of these results
to predict the distribution of excesses at 5Myr from measurements of the distribution at 10 - 30 Myr, where
enough systems have been measured to define the distribution well. We use three samples (Sco-Cen, Orion Ob1, and 
R05), each of which includes the complete range of [24]$_{obs}$-[24]$_*$ down to zero, i.e.,
photospheric colors.  If the scaled colors for sources predict much larger excesses than the 5Myr old
Orion Ob1b excesses, then we can conclude that the mid-IR colors of our samples from 5 to 15 Myr do not
follow a $t^{-1}$ decline.

Figure \ref{scaled} shows the scaled Sco-Cen and Rieke et al. excesses compared to the observed 5 Myr excesses in Orion Ob1b 
normalized to the total number of sources in each sample.
Many scaled excess sources ($\sim$ 20\% of the total population) are $\gtrsim$ 1-3 mags redder than any in Orion Ob1b.  
The Kolmogorov-Smirnov test shows that the scaled Sco-Cen (Rieke et al.) sources  
have a probability of 0.07 (0.06) of being drawn from the same population as Orion Ob1b with a $\gtrsim$ 0.75 mag excess.
If we instead compare the scaled sources to Orion Ob1b sources with a $\gtrsim$ 0.5 mag excess, where Orion and the 
other two populations begin to overlap in Figure \ref{scaled}, the probability is even lower:
 2.3$\times$10$^{-5}$ (3.5$\times$10$^{-8}$) for the Sco-Cen (Rieke et al.) sample.  Thus, the evolution of 24 $\mu m$ 
emission from Orion Ob1b, Sco-Cen, and Rieke et al. sources is not consistent with a t$^{-1}$ decay.

There are two main alternatives to the t$^{-1}$ decay of IR excess with time for the youngest debris disks.  
The IR excess could be constant for $\sim$ 20 Myr and then follow a t$^{-1}$ decay law.  The IR excess could also
increase with time to some peak value and then follow a t$^{-1}$ decay law.  To test the constant emission 
possibility, we use the Wilcoxon Rank-Sum test.  The Rank-Sum test allows us to 
evaluate whether or not the populations have the same mean value or have 
intrinsically larger/smaller excesses than another sample (Z parameter). 
The test also measures the probability that two samples are drawn from 
the same parent population by the Prob(RS) parameter (as in the K-S test).

Table 5 summarizes our results.
Statistical tests show that the [24]$_{obs}$-[24]$_{\star}$ excesses cannot be constant with time and verify that 
emission rises from 5 to 10 Myr (Table 3).  Sco-Cen has the largest mean excess ratio ($\sim$ 1.25) and has  
 the widest range of colors with a [24]-[24]$_{\star}$ standard deviation of $\sim$ 1.4 compared 
to $\sim$ 0.6-0.8 for the Orion subgroups.   The Wilcoxan Rank-Sum test reveals that Sco-Cen has $\lesssim$ 5\% probability of being drawn 
from the same population as Orion Ob1b and has a statistically significant larger peak (Z $\sim$ -1.65).  Orion Ob1a also 
has a significant larger peak than Orion Ob1b (Z $\sim$ -2.8, Prob (RS) $\sim$ 0.002), and Sco-Cen's peak is marginally larger 
than Orion Ob1a's (Z $\sim$ -0.05).  

These results lead us to conclude that \textit{\textbf{the evolution of mid-IR excess emission from planet formation 
in debris disks is best characterized as a rise in excess  
from $\sim$ 5-10 Myr, a peak at $\sim$ 10-15 Myr, and a fall in excess from $\sim$ 15/20 Myr-1 Gyr}}.
The rise to maximum excess from 5 to 10 Myr is steep: typical excesses increase from $\sim$ 1 mag to $\sim$ 3 mag by 11.8 Myr.  
The peak in excess amplitude is $\sim$ 5-10 Myr broad because
the excesses in NGC 7160, h and $\chi$ Per, and Sco-Cen are all comparable.  By $\sim$ 25 Myr the typical excesses decline 
 with age as suggested by R05.   

\subsection{Comparison with Models of Emission from Planet Formation}
To investigate how the 'rise and fall' trend of debris emission may be connected with physical processes producing 
the emission, we overplot the debris disk evolution tracks (dotted lines) from \S 3.3.3 and debris disk tracks for a low-mass disk 
(1/3$\times$MMSN$_{scaled}$).   The KB04 calculations start at t=0 with an ensemble of $\sim$ km-sized planetesimals.
To bracket the likely timescale for km-sized planetesimals to form at 30-150 AU (e.g. Dominik \& Dullemond 2005; Weidenschilling 1997), 
we include a second locus shifted by 2 Myr (solid lines).  The range in debris disk masses 
is about a factor of 10, comparable to the range of disk masses inferred from submillimeter observations of young stars
\citep{Aw05}.

The debris disk models from \S 3.3.3 show a steep 
increase in debris emission from 5-10 Myr, a plateau for the following $\sim$ 20 Myr, and then a shallow decline in debris emission.
The massive disk locus (dotted line) yields a peak in emission at $\sim$ 7-8 Myr, or very close to the age of HR 4796A.  
The locus started at 2 Myr (solid line) peaks at $\sim$ 9-10 Myr, close to the ages of h and $\chi$ Persei and NGC 7160, 
and yields substantial emission through 20 Myr before emission declines.  The lower-mass disk loci
peak later at $\sim$ 40 Myr with excesses comparable to the majority of those in R05.  Caution should be taken to avoid 
overinterpreting these similarities: the exact time of the debris disk emission peak as 
well as the amplitude of the peak depend on 
input parameters such as planetesimal disruption energy.  
Nevertheless, the massive debris disk model yields the same 
general trend in the maximum 24 $\mu m$ excess amplitude with time;
the low-mass disk model reproduces the 24 $\mu m$ excesses of many $\gtrsim$ 30 Myr-old sources.
The observed behavior of 24 $\mu m$ excess with time is then at least qualitatively
consistent with our understanding of the processes associated with planet formation.
\section{Discussion}
\subsection{Summary of Results}
Our analysis of MIPS data for the 13-14 Myr-old double cluster, 
h and $\chi$ Persei, shows two significant 24 $\mu m$ excess populations.  
Bright Be stars with J$\lesssim$ 12-13 have 1-2 mag excesses 
at 24 $\mu m$ and follow a clear Be star locus in the J-H/H-K$_{s}$ color-color diagram.  Optical 
spectra confirm the Be star status for just under half of the candidates from the color-color diagram.
We also detect a B4 star with a clear 24 $\mu m$ excess but without H$_{\alpha}$ emission 
or evidence for near-IR excess.

Fainter stars with J $\sim$ 14-15 fall on the 14 Myr isochrone in a J/J-H color-magnitude diagram.  
Optical spectra confirm that many of these stars have late A-type or F-type spectra, consistent with cluster 
membership.  The IRAC and MIPS colors of these sources suggest that the frequency of excess 
at wavelengths which probe IR excess emission increases with increasing wavelength.  The wavelength-dependent 
frequency of excess is consistent with the presence of inner holes devoid of dust.

Our analysis of the dust temperatures in the fainter excess sources suggest two groups.  
A smaller group of stars has emission from warmer dust with T $\sim$ 200-300 K.  A larger 
group has emission from colder dust, T $\lesssim$ 200 K.  In both groups, the dust luminosity is 
a small fraction of the stellar luminosity, L$_{d}$/L$_{\star}$ $\sim$ 10$^{-4}$-10$^{-3}$, typical 
of debris disks like HR 4796A (Low et al. 2005).  The IR colors and spectral energy distributions of the latter 
group are consistent with predictions from cold debris disk models; sources with warmer dust 
may have terrestrial zone debris emission (see also C07b). 

The MIPS data from h \& $\chi$ Persei and other recently surveyed clusters yield a large sample 
of disks at 5-20 Myr, an age range critically
important for understanding debris disk evolution and planet formation. 
This sample shows that debris disk emission rises from 5 Myr to $\sim$ 10 Myr, peaks at $\sim$ 10-15 Myr, 
 and then fades on a $\sim$ 150 Myr timescale as t$^{-1}$ (R05).  Numerous statistical tests verify 
the observed trend.
Debris production from ongoing planet formation explains the general time evolution of this emission 
(e.g. KB04).  

\subsection{Future Modeling Work: Explaining the Range of MIPS Excesses}
The debris disk models from Kenyon \& Bromley (2004a) generally explain the peak excesses 
for sources in h \& $\chi$ Per and other $\sim$ 10-15 Myr-old clusters.
 However, at a given age stars have a wide spread of IR excesses above and below the debris disk 
model predictions.  The IR excesses far weaker than the model predictions have several tenable explanations.  Low-mass disks 
modeled in \S 4 have weaker excesses.  
Disks in systems with binary companions close to the disk radius are probably disrupted quickly, although
binaries with wider separations should have little effect, and very close separations may actually enhance
infrared excesses (Bouwman et al. 2006; Trilling et al. 2007).  Gas giant planets may also remove IR-emitting dust.  

Reproducing the larger IR excesses (K$_{s}$-[24] $\gtrsim$ 4-5) is more difficult.  
The debris disk model used in this paper yields a
peak K$_{s}$-[24] $\sim$ 3.5 (unreddened), but HR 4796A and several sources in h and $\chi$ Per and Sco-Cen 
have stronger excesses. 
More massive disks should yield stronger 24$\mu m$ excesses, but the disk mass cannot be increased indefinitely.  
A disk with mass M$_{d}$ $\gtrsim$ 0.1-0.15 M$_{\star}$ would have been initially gravitationally 
unstable and would form a low-mass companion.  
However, the debris disk status of one of these extreme cases, HR 4796A with K$_{s}$-[24] $\approx$ 5, 
has been confirmed by extensive disk SED modeling (e.g. \citealt{Au99, Cu03, Wa05})
and strict gas mass upper limits of $\lesssim$ 1 M$_{Jupiter}$ \citep{Ck04}.   
There are several ways to account for these larger excesses.  
For example, dynamical processes that allow small grains to be retained (which produce 
larger opacity) in rings like that observed for HR 4796A may explain the large-amplitude excesses in some debris disks 
 (e.g. Klahr and Lin 2005; Takeuchi and Artymowicz 2001).
 Whether or not the strong excess sources in h and $\chi$ Per can be explained by such grain confinement 
 mechanisms is the subject of future work.
\subsection{Comparison with Previous Spitzer Observations and Analysis of h and $\chi$ Persei  
from \citet{Cu07a} and \citet{Cu07b}}
This paper completes the first study of the circumstellar disk population of pre-main sequence 
stars in the massive double cluster, h and $\chi$ Persei.  Together with 
Cu07a and Cu07b, this work provides new constraints on the frequency, lifetimes, 
and evolutionary states of circumstellar disks in 10-15 Myr old stars.  
Here we summarize the main results and conclusion of these studies.

Spitzer data for h and $\chi$ Per provide clear
evidence that the frequency of IR excess emission
depends on wavelength and on the mass of the star
(C07a; this paper, \S 3). Stars in both clusters
have a higher frequency of IR excess at longer
wavelengths. Lower mass (1.4-2 M$_{\odot}$) stars have IR excesses more
often than more massive ($\gtrsim$ 2 M$_{\odot}$) stars. Su et al. (2006)
and \citet{Gr07} derive similar results
for other clusters. Taken together, these results
are consistent with an inside-out clearing of
dust from young circumstellar disks, as expected
from theoretical models of planet formation (e.g. \citealt{KB04b}).

To compare the completeness level of the MIPS sample with
the IRAC sample from C07a, we derive the fraction of IRAC
sources with MIPS detections at each IRAC band. The IRAC
survey has 90\% completeness levels of $\sim$ 14.5 at
 [4.5] and $\sim$ 13.75 at [8]. The
MIPS sample includes 87\% (3\%) of the IRAC sources with
[3.6] $<$ 10 ([3.6] $<$ 14.5) and 88\% (11\%) of the IRAC sources
with [8] $<$ 10 ([8] $<$ 13.75) within 25' of either cluster center.
 Because the MIPS survey detects
such a small percentage of the IRAC sources in C07a, we
cannot analyze a statistically significant population of
IRAC IR excess sources with MIPS detections. However, the
MIPS sources yield interesting constraints on the Be star
population in h and $\chi$ Per (\S 3.2) and demonstrate a
clear peak in the time evolution of the 24 $\mu$m excess of
debris disks (\S 4.).

Detailed analyses of the IRAC/MIPS colors and the
broadband SEDs demonstrate that warm dust (T $\sim$
240--400 K) is visible in 11 cluster stars (C07b;
this paper, \S 3). The dust luminosities of ten of these
sources ($\sim$ 10$^{-4}$-6$\times$10$^{-3}$ L$_{\star}$) suggest this emission arises
from optically thin dust in a debris disk. The
IR excesses of these sources -- which comprise the
majority of known warm debris disks (see Hines et al. 2006, Wyatt et al. 2007b, 
and \citealt{Gr07} for others) -- are
consistent with detailed calculations of terrestrial
planet formation around $\sim$ 2 $M_{\odot}$ stars \citep{KB04a}.

Most of the IRAC/MIPS IR excess sources show evidence for
cooler dust with T $\sim$ 100--200 K (this paper; \S 3, 4).
Although the lack of 70 $\mu$m detections prevents
us from deriving precise limits on the dust temperatures
and luminosities, the Spitzer data suggest that most
(perhaps all) of these sources are debris disks with
SEDs similar to those observed in Sco-Cen, 
the TW Hya Association and other young clusters. (\citealt{Ch05b, Lo05}).
When combined with data from the literature, these
data provide clear evidence for a rise in the magnitude
of the IR excesses from debris disks from $\sim$ 5 Myr
to $\sim$ 10--15 Myr followed by a fall from $\sim$
20 Myr onward.

Although theory provides a reasonably good first-order
explanation for the time evolution of the IR excesses in
$\gtrsim$ 1.4 M$_{\odot}$ stars,
some aspects of the observations remain challenging.
A large range of initial disk masses and binary companions
can probably explain the large range in IR excesses at
a given stellar age, but these explanations require
further testing.
Current theory does not explain the largest IR excesses
observed in the 10--20 Myr old stars in h and $\chi$ Persei, 
Sco-Cen, and the TW Hya Association (specifically HR 4796A). 
Dynamical, radiative, and stochastic processes not included
in the numerical calculations are possible solutions to
this failure. Increasing the sample size of this extreme
population would provide better constraints on these
processes.

 Finally, this survey has provided us with several interesting
sources that warrant more detailed investigation. 
For instance, 
'Source 5' -- discussed here and in \citet{Cu07b} -- is probably an extremely massive debris disk.  
With a fractional disk luminosity of $\sim$ 6$\times$ 10$^{-3}$, its emission rivals that of  
 HR 4796A, $\beta$ Pic and other massive, nearby debris disks.  However, this source differs from these other 
sources in at least two important ways.
First, its spectral type is later (F9) than most stars with massive debris disks.  Second, it harbors far warmer 
dust (T$_{d}$ $\sim$ 240-330 K) than most massive debris disks like HR 4796A (T$_{d}$ $\sim$ 110 K; \citealt{Lo05}).  This 
feature may make it more similar to the warm debris disk of HD 113766A in Sco-Cen, the second most 
luminous source in Sco-Cen shown in Figure \ref{excvage} \citep{Ch05b}, than to HR 4796A and $\beta$ Pic. 
Longer wavelength observations of this h and $\chi$ Per source (e.g. 30-100 $\mu m$) will better constrain its SED and thus 
its dust population(s).  Mid-IR spectroscopy of this source may also provide clues to the chemical composition of its 
circumstellar dust to compare with models of cometary and asteroidal material.
\subsection{Future Observations}
Future observations of h and $\chi$ Persei will provide stronger constraints on debris disk evolution and 
 the possibilities for producing the wide range of debris disk emission.  A deeper MIPS survey (approved for Spitzer cycle 4)
of the double cluster will identify $\gtrsim$ 1000-2000 cluster sources with [24] $\lesssim$ 12.25, 
the brightness of an 1.3 M$_{\odot}$ G9 (1.7 M$_{\odot}$ A8) star with a 3 (2) magnitude excess.  
This $\sim$ 2 magnitude increase in sensitivity ($\sim$ 2260 seconds/pixel integration) 
should yield a larger sample of 24 $\mu m$ excess sources which will better map out the 
distribution of mid-IR excesses during the primordial-to-debris disk transition.  
If the correlation of massive, high fractional luminosity-disks with early A stars is purely a selection 
effect, then this deeper survey of h and $\chi$ Persei should reveal many massive debris disks around 
slightly later-type stars like 'Source 5'.  This survey will be complemented by a deeper IRAC survey (also approved for cycle 4) 
of the double cluster, which will identify h and $\chi$ Per sources with [5.8] ([8]) $\lesssim$ 15.9 (15.2), the 
brightess of a $\sim$ 0.8 M$_{\odot}$ M0 (1.0 M$_{\odot}$ K6) photosphere.  The $\sim$ 1.5-2 magnitude increase in sensitivity 
($\sim$ 120 seconds/pixel integration vs. 20.8 seconds/pixel from the C07a observations) will likely result in photometry for 
$\gtrsim$ 10,000-15,000 cluster stars through [8], assuming a typical cluster initial mass function (e.g. Miller and Scalo 1979).
These two surveys will likely detect hundreds of debris disk (and perhaps transition disk) candidates 
and yield extremely strong constraints on evolution of dust in circumstellar disks 
from warm, inner regions (IRAC) to cooler regions (MIPS) at a critical age for planet formation. 

Ground-based surveys of h and $\chi$ Persei may also provide important clues about the evolution of 
disks around young stars.  For instance, the ability of binary companions to 
affect the mid-IR excesses from disks can also be tested.  At $\sim$ 2.34 kpc, a binary system with separation of 
$\sim$ 100 AU (and thus able to truncate debris disks) has an angular separation of $\approx$ 0.04".  Such systems
can be resolved by long-baseline interferometers such as the Keck Interferometer.  Comparing the IR excesses from 
single and binary systems can then determine if weaker IR excess sources are binaries.  
A large-scale spectroscopic survey of all sources in h \& $\chi$ Per brighter than V $\sim$ 21 ($\gtrsim$ 10,000) 
is underway (Currie et al. 2007, in prep.).  This survey will identify sources most likely to be h \& $\chi$ Per members 
as well as those with strong H$_{\alpha}$ emission indicate of gas accretion.  Preliminary work indicates that 
the population of accreting h and $\chi$ Per sources is non negligible ($\gtrsim$ 20-30; \citealt{Cu07c}).  Comparing the IR excesses 
of accreting sources with those that are not accreting may examine the role of residual circumstellar gas in 
affecting the mid-IR excesses from disks.
\acknowledgements
We thank the referee for a thorough review and suggestions which improved this manuscript.
We also thank Michael Meyer, John Carpenter, Christine Chen, and Nadya Gorlova for useful discussions regarding debris disks 
in other clusters; Matt Ashby, Rob Gutermuth, and Anil Seth provided  
 valuable advice regarding galaxy contamination.  We acknowledge from the 
NASA Astrophysics Theory Program grant NAG5-13278, TPF Grant NNG06GH25G, and the Spitzer GO program (Proposal 20132).  
T.C. is supported by a SAO predoctoral fellowship; Z.B. received support from Hungarian OTKA Grants 
TS049872 and T049082.  This work was partially supported by contract 1255094, issued by JPL/Caltech to the University 
of Arizona.

\clearpage
\begin{deluxetable}{llllllllllllllllll}
\tiny
\setlength{\tabcolsep}{0.01in}
\tabletypesize{\tiny}
\tablecolumns{18}
\tablewidth{0pt}
\tablecaption{MIPS sources with 2MASS/IRAC counterparts in h \& $\chi$ Persei}
\tiny
\tablehead{{$\alpha$}&{$\delta$}&{V}&{B-V}&{U-B}&{J}&{H}&{$K_{s}$}&{[3.6]}&{[4.5]}&{[5.8]}&{[8]}&{\textbf{[24]}}&{$\sigma$([3.6])}&{$\sigma$([4.5])}&{$\sigma$([5.8])}&{$\sigma$([8])}&{\textbf{$\sigma$([24])}}}
\startdata
35.4810&57.2429& 6.480& 0.502&-0.050&5.08&4.86&4.77&8.39&\ldots&4.73&4.52&4.14&0.01&0.00&0.00&0.00&0.03\\
34.8081&57.1693& 6.700& 0.503&-0.428&5.53&5.37&5.31&\ldots&\ldots&5.30&5.07&4.64&0.00&0.00&0.00&0.00&0.03\\
34.7685&57.1355& 6.567& 0.452&-0.346&5.56&5.43&5.29&8.18& 7.05&5.25&5.17&4.68& 0.01&0.00&0.00&0.00&0.00 \\
35.7517&57.3870& 6.977& 0.707&-0.249&5.59&5.38&5.26&8.31&\ldots&5.16&5.04&4.73& 0.04&0.00&0.00&0.00&0.03\\
34.2155&57.0552&\ldots&\ldots&\ldots&5.78&5.71&5.61&\ldots&\ldots&\ldots&\ldots&5.00&0.00&0.00&\ldots&\ldots&0.03\\
\enddata
\tablecomments{First five entries in our photometry catalogue from MIPS, 2MASS and IRAC.  We also include optical UBV photometry 
from Slesnick et al. (2002).  Errors in the 
2MASS JH$K_{s}$ filters are $\lesssim$ 0.05 for sources except the $\sim$ 21 fainter (J$\sim$ 14-15) sources.
}
\end{deluxetable}

\begin{deluxetable}{lllllllllll}
\tiny
\setlength{\tabcolsep}{0.03in}
\tabletypesize{\tiny}
\tablecolumns{11}
\tablewidth{0pt}
\tablecaption{Be stars/candidates and selected evolved stars with 24 $\mu m$ excess in h \& $\chi$ Persei}
\tiny
\tablehead{{Number}&{Oos. Num.}&{Spec. Type}&{$\alpha$}&{$\delta$}&{V}&{J}&{H}&{$K_{s}$}&{[8]}&{[24]}}
\startdata
1& 2589  &   Be  &  35.7229  &  57.2452  &  7.48  &  5.89  &  5.62  &  5.47  &  4.92  &  4.50 \\
2&  847  &   B1.5Ie  &  34.6997  &  57.0673  &  9.13  &  8.30  &  8.19  &  8.10  &  7.61  &  6.42 \\
3& 1261  &   B3Ve  &  34.8605  &  57.0784  &  9.56  &  8.40  &  8.22  &  7.87  &  6.30  &  4.93 \\
4& 2284  &   B0e  &  35.5268  &  57.0903  &  9.68  &  8.48  &  8.10  &  7.55  &  5.65  &  4.76 \\
5& 2402  &   B1e  &  35.5947  &  57.2847  &  9.64  &  8.50  &  8.38  &  8.33  &  7.02  &  6.67 \\
6& 2088  &   B1IIIe  &  35.4308  &  57.1258  &  9.49  &  8.82  &  8.67  &  8.58  &  8.43  &  6.30 \\
7& 1161  &   B1Ve  &  34.8069  &  57.1288  & 10.22  &  9.07  &  8.91  &  8.68  &  7.60  &  6.42 \\
8&  846  &   B1e  &  34.7015  &  57.2403  &  9.98  &  9.08  &  9.10  &  9.06  &  8.42  &  7.49 \\
9&  517  &   B3e  &  34.5629  &  57.1711  & 14.59  &  9.30  &  9.22  &  9.08  &  8.53  &  7.43 \\
10& 2165  &   B1Ve  &  35.4705  &  57.1664  & 10.15  &  9.47  &  9.42  &  9.41  &  7.81  &  6.76 \\
11& 2566  &   B1e  &  35.7004  &  57.2002  & 10.63  &  9.58  &  9.46  &  9.30  &  9.14  &  8.16 \\
12& 1282  &   09e  &  34.8701  &  57.1179  & 11.00  &  9.64  &  9.45  &  9.21  &  7.79  &  6.85 \\
13& 2649  &   B2e  &  35.7674  &  57.1275  & 10.64  &  9.68  &  9.59  &  9.49  &  9.07  &  8.32 \\
14& 2242  &   B3e  &  35.5103  &  57.1557  & 10.96  & 10.10  &  9.91  &  9.67  &  9.05  &  7.85 \\
15& 1438  &   B2e  &  34.9493  &  57.1110  & 11.20  & 10.18  & 10.02  &  9.87  &  8.47  &  7.62 \\
16& 1278  &   B2e  &  34.8703  &  57.1901  & 11.59  & 10.66  & 10.49  & 10.29  &  9.50  &  8.00 \\
17& 2091  &   B2e  &  35.4353  &  57.1812  & 11.72  & 10.69  & 10.49  & 10.30  &  9.17  &  7.86 \\
18& 1114  &   B3e  &  34.7859  &  57.0636  & 12.40  & 11.19  & 10.90  & 10.69  &  9.48  &  7.88 \\
19& 1977  &   B2e  &  35.3538  &  57.1979  & 12.28  & 11.23  & 11.03  & 10.82  &  9.66  &  8.21 \\
20&  563  &   Be  &  34.5822  &  57.1462  & 12.27  & 11.53  & 11.47  & 11.45  & 11.30  & 10.38 \\
21&   99  &   B3I  &  34.8081  &  57.1693  &  6.7  &   5.53  &   5.37  &   5.31  &   5.07  &   4.64 \\
22&   99  &   99  &  34.2405  &  57.1302  &  99.00  &   8.09  &   7.89  &   7.52  &   6.32  &   5.24 \\
23&   99  &   B1Ie  &  35.4617  &  57.3866  &  9.27  &   8.36  &   8.21  &   8.08  &  99.00  &   5.30 \\
24&   99  &   99  &  35.3253  &  57.3062  &  9.74  &   8.57  &   8.35  &   8.11  &   6.97  &   5.56 \\
25&   99  &   99  &  35.1627  &  57.3119  &  9.64  &   8.93  &   8.80  &   8.67  &   8.83  &   7.45 \\
26&   99  &   99  &  35.8538  &  57.3177  &  10.60  &   9.02  &   8.78  &   8.53  &   7.33  &   5.95 \\
27&   99  &   99  &  34.5765  &  56.8506  &  9.70  &   9.30  &   9.29  &   9.27  &   9.33  &   8.30 \\
28&   99  &   99  &  35.8877  &  57.0757  &  10.45  &   9.34  &   9.09  &   8.81  &   8.62  &   6.86 \\
29&   99  &   99  &  33.9481  &  57.4207  &  99.00  &   9.47  &   9.39  &   9.21  &  99.00  &   7.44 \\
30&   99  &   99  &  34.1349  &  57.5337  &  99.00  &   9.48  &   9.40  &   9.31  &  99.00  &   8.39 \\
31&   99  &   99  &  35.6957  &  56.9683  &  10.70  &   9.54  &   9.24  &   8.91  &   7.72  &   6.29 \\
32&   99  &   99  &  35.7228  &  56.7871  &  10.61  &   9.87  &   9.81  &   9.75  &   9.82  &   9.08 \\
33&   99  &   99  &  35.8673  &  57.3905  &  11.19  &   9.91  &   9.72  &   9.48  &   8.34  &   6.97 \\
34&   99  &   99  &  36.1358  &  57.0125  &  99.00  &   9.94  &   9.71  &   9.40  &  99.00  &   6.88 \\
35&   99  &   99  &  35.2199  &  57.2244  &  11.62  &  10.17  &   9.90  &   9.62  &   9.51  &   7.70 \\
36&   99  &   A2  &  35.6384  &  57.0416  &  10.79  &  10.24  &  10.19  &  10.12  &  10.08  &   9.17 \\
37&   99  &   99  &  35.9502  &  56.7838  &  11.27  &  10.31  &  10.26  &  10.22  &  10.14  &   9.52 \\
38&   99  &   99  &  36.1420  &  56.8837  &  99.00  &  10.36  &  10.17  &  10.15  &  99.00  &   7.30 \\
39&   99  &   99  &  36.0424  &  57.3254  &  11.85  &  10.40  &  10.23  &  10.21  &  99.00  &   8.97 \\
40&   99  &   99  &  35.0563  &  56.7401  &  99.00  &  10.42  &  10.40  &  10.38  &   8.64  &   7.81 \\
41&   99  &   99  &  34.4142  &  57.0958  &  11.25  &  10.51  &  10.45  &  10.40  &  10.33  &   9.16 \\
42&   99  &   99  &  34.9078  &  56.7302  &  11.40  &  10.69  &  10.58  &  10.56  &  10.57  &   9.90 \\
43&   99  &   99  &  34.3799  &  56.9881  &  11.27  &  10.70  &  10.70  &  10.65  &  10.65  &   9.96 \\
44&   99  &   99  &  33.9740  &  57.1986  &  99.00  &  10.96  &  10.58  &  10.47  &  99.00  &   9.71 \\
45&   99  &   99  &  35.0169  &  57.0555  &  11.91  &  11.07  &  10.97  &  10.92  &  10.90  &  10.07 \\
46&   99  &   99  &  34.9094  &  56.8205  &  11.62  &  11.08  &  11.09  &  11.06  &  11.11  &  10.08 \\
47&   99  &   99  &  35.8650  &  57.3698  &  12.17  &  11.09  &  10.89  &  10.85  &  10.64  &   9.61 \\
48&   99  &   F7  &  35.7907  &  57.1709  &  12.39  &  11.12  &  10.84  &  10.80  &  10.77  &  10.03 \\
49&   99  &   99  &  34.5332  &  57.3827  &  13.13  &  11.53  &  11.31  &  11.17  &  11.11  &  10.47 \\
50&   99  &   99  &  36.3024  &  57.0817  &  99.00  &  11.63  &  11.38  &  11.30  &  99.00  &  10.54 \\
51&   99  &   G2  &  34.6765  &  57.2288  &  13.18  &  11.66  &  11.32  &  11.17  &  11.15  &  10.45 \\
52&   99  &   G1  &  35.4931  &  57.1820  &  13.26  &  11.67  &  11.28  &  11.18  &  11.10  &  10.49 \\
53&   99  &   B4V &  34.5567  &  57.2121  &  12.46  &  11.71  &  11.71  &  11.62  &  11.68  &  10.57 \\
54&   99  &   99  &  34.4228  &  57.4477  &  13.34  &  11.75  &  11.36  &  11.34  &  11.29  &  10.68 \\
55&   99  &   99  &  34.7373  &  57.4550  &  13.31  &  12.02  &  11.84  &  11.79  &  11.69  &  10.72 \\
56&   99  &   99  &  35.5442  &  57.5230  &  13.40  &  12.14  &  11.96  &  11.87  &  11.59  &  10.30 \\
57&   99  &   99  &  35.3152  &  57.2595  &  13.91  &  12.84  &  12.71  &  12.59  &  12.32  &  10.63 \\
\enddata
\tablecomments{List of Be stars, Be star candidates, and selected evolved stars with 24 $\mu m$ 
excess emission in h and $\chi$ Persei.  We include the Oosterhoff (1937) 
number, spectral type, luminosity class (I-supergiant, III-giant, V-dwarf), and V magnitude where available.  
All sources with the 'e' designiation in spectral type are Be stars.  The first twenty entries are 
confirmed Be stars from Bragg and Kenyon (2002) and have Oosterhoff numbers.  Slesnick et al. (2002)
identify an additional Be star in lower density regions surrounding the center of h and $\chi$ Per (source 
number 23).  All the other stars (referred to as 'candidate Be stars' in the text) 
are either supergiants, foreground stars, or 
B-type cluster members from either Slesnick et al. (source 21) or archived spectroscopic data from FAST (the 
five other sources).}
\end{deluxetable}

\begin{deluxetable}{lllllllllllllllllll}
\tiny
\setlength{\tabcolsep}{0.02in}
\tabletypesize{\scriptsize}
\tablecolumns{9}
\tablewidth{0pt}
\tablecaption{Observed Properties of faint MIPS excess sources in h \& $\chi$ Persei consistent with membership}
\tiny
\tablehead{{$\alpha$}&{$\delta$}&{ST}&{J}&{$K_{s}$}&{K$_{s}$-[4.5]}&{K$_{s}$-[5.8]}&{K$_{s}$-[8]}&{K$_{s}$-[24]}}
\startdata
35.0411&57.1494&A2&13.84&13.62& 0.16&0.27&-0.26&2.81\\
35.4339&57.0355&A2?&13.92&13.71& 0.17&0.21&0.21&3.14\\
36.0559&57.1190&A2?&13.96&13.45& 99&99&99&2.36\\
35.3783&57.2056&A2?&14.07&13.81& 0.22&0.31&0.24&3.02\\
35.1178&57.2625&A3?&14.10&13.83& 0.15&0.18&0.06&3.17\\
34.7041&56.9259&A4?&14.17&13.69& 0.18&0.09&0.19&4.38\\
36.2329&57.0037&A6?&14.30&14.01& 99&99&99&3.86\\
35.3346&57.5054&A6&14.39&14.28& 0.35&-0.27&99&3.53\\
36.0860&57.0493&A6?&14.43&13.76& 99&99&99&4.18\\
34.6889&57.2731&A6&14.44&13.99& 0.13&0.33&0.43&3.67\\
34.8108&57.4067&F2?&14.59&14.00& 0.21&0.26&0.26&3.67\\
35.5144&57.1202&F2&14.66&14.20& 99&99&99&3.16\\
35.3215&57.2030&F9&14.67&14.13& 0.25&0.33&0.47&3.26\\
35.1171&57.2662&F9?&14.77&14.27& 0.28&0.56&0.64&3.47\\
35.4562&57.2245&F3&14.87&14.36& 0.11&99&99&3.63\\
34.7824&57.2347&F9&15.10&14.47& 0.30&0.56&1.37&4.55\\
34.9448&57.1925&F9?&15.12&14.51& 0.45&0.90&1.48&5.65\\
\enddata
\tablecomments{Faint MIPS excess sources that are consistent with membership in h \& $\chi$ Persei.  Spectral types 
for seven sources are derived from Hectospec data while the rest were inferred from J-band photometry.  
Values of 99 denote sources without 5$\sigma$ detections in a given band.  Magnitudes 
given are the observed, not dereddened, values.}
\end{deluxetable}

\begin{deluxetable}{lllllllllllllllllll}
\tiny
\setlength{\tabcolsep}{0.02in}
\tabletypesize{\scriptsize}
\tablecolumns{9}
\tablewidth{0pt}
\tablecaption{Inferred Properties of faint MIPS excess sources in h \& $\chi$ Persei consistent with membership}
\tiny
\tablehead{{$\alpha$}&{$\delta$}&{ST}&{T$_{d}$}&{T$_{d}$}&{$\chi$$^{2}$}&{L$_{\star}$}&{L$_{d}$/L$_{\star}$}&{R$_{d}$}&{Disk}\\
{}&{}&{}&{FR}&{BB}&{}&{(L$_{\odot}$)}&{}&{(AU)}&{Type}}
\startdata
35.4339&57.0355&A2?&185&185&1.7&18.1&7$\times$10$^{-4}$&9.75&CDD\\
35.3783&57.2056&A2?&170&175&5.7&18.1&5.5$\times$10$^{-4}$&10.9&CDD\\
35.1178&57.2625&A3?&100&90&2.0&15.7&1.9$\times$10$^{-3}$&38.4&CDD\\
34.7041&56.9259&A4?&130&120&6.0&12.8&3.5$\times$10$^{-3}$&19.5&CDD\\
34.6889&57.2731&A6&230&85,375&7.6&12&3.1$\times$10$^{-3}$&1.8,37.0&WDD\\
34.8108&57.4067&F2?&170&165&3.1&6.7&2.3$\times$10$^{-3}$&7.5&CDD\\
35.3215&57.2030&F9&240&220,305&3.2&6.1&1.9$\times$10$^{-3}$&2.1,4&WDD\\
35.1171&57.2662&F9?&250&175,400&5.8&6.1&2.3$\times$10$^{-3}$&1.2,6.3&WDD\\
34.7824&57.2347&F9&300&240,330&0.5&6.1&6.3$\times$10$^{-3}$&1.8,3.4&WDD\\
34.9448&57.1925&F9?&230&110,435&5.1&6.1&1.5$\times$10$^{-2}$&1.0,16.0&TWH\\
\enddata
\tablecomments{Inferred properties of faint MIPS excess sources with photometry at multiple bands.
The SEDs of 10 sources were constrained well enough 
to derive disk temperatures from flux-ratio diagrams (T$_{D}$ FR) and from fitting the source SEDs to star + 
one (two) blackbody disk populations (T$_{D}$ BB) for sources without (with) IRAC excess.  The location of the dust (R$_{D}$) 
is derived from simple blackbody equilibrium. Relative disk luminosities (L$_{D}$/L$_{\star}$) 
were derived assuming a stellar luminosity from stars of a given spectral type at the age of h and $\chi$ Per 
from Siess et al. (2000).  For the evolutionary states, WDD = warm debris disk (which have colder components), 
CDD = cold debris disk, and TWH = A 'TW Hya'-like source that may 
be optically-thick at long wavelengths.}
\end{deluxetable}

\begin{deluxetable}{lllllllll}
\tiny
\tabletypesize{\tiny}
\tablecolumns{7}
\tablecaption{Statistical tests for complete samples of MIPS-detected stars}
\tiny
\tablehead{{Group}&{Age}&{$\bar{[24]-[24]_{\star}}$(All)} & {$\bar{[24]-[24]_{\star}}$(Excess)} & 
{$\sigma$([24]-[24]$_{\star}$} & {R-S Z (Excess)} & {R-S Prob (Excess)}}
\startdata
Sco-Cen & 16 Myr & 0.67 & 2.10 & 1.41 &  -1.65, -0.05 & 0.05, 0.48\\
Orion Ob1a & 10 Myr & 0.72 & 1.46& 0.83 & -2.8, 0 & 0.002, 1\\
Orion Ob1b & 5 Myr & 0.36 & 0.91 & 0.63 & 0, 2.8 & 1, 0.002
\enddata
\tablecomments{Statistics comparing the populations of Sco-Cen, 
Orion Ob1a, and Orion Ob1b.  We calculate the 
mean ($\bar{[24]-[24]_{\star}}$) color (for the entire sample and for 'excess' sources), 
the standard deviation ($\sigma$) of each sample's colors, and the Wilcoxan Rank-Sum probability 
and Z parameter.  Here 'excess' sources denote those with 
[24]-[24]$_{\star}$ $\gtrsim$ 0.25.   
The first (second) entry in the R-S test statistics compares each population to 
Orion Ob1b (Orion Ob1a).  A positive Z parameter means that the sample has a larger peak value than Orion Ob1b 
(first entry) and Orion Ob1a (second entry).  A low R-S probability means that the [24]-[24]$_{\star}$ colors of the 
populations are very different.  These results show that Sco-Cen and Orion Ob1a have statistically significant 
larger 24$\mu m$ excesses than Orion Ob1b.  Sco-Cen and Orion Ob1a have similar populations, though Sco-Cen has 
larger mean excesses and a larger dispersion in excesses.}
\end{deluxetable}

\clearpage
\begin{figure}
\epsscale{1.}
\plottwo{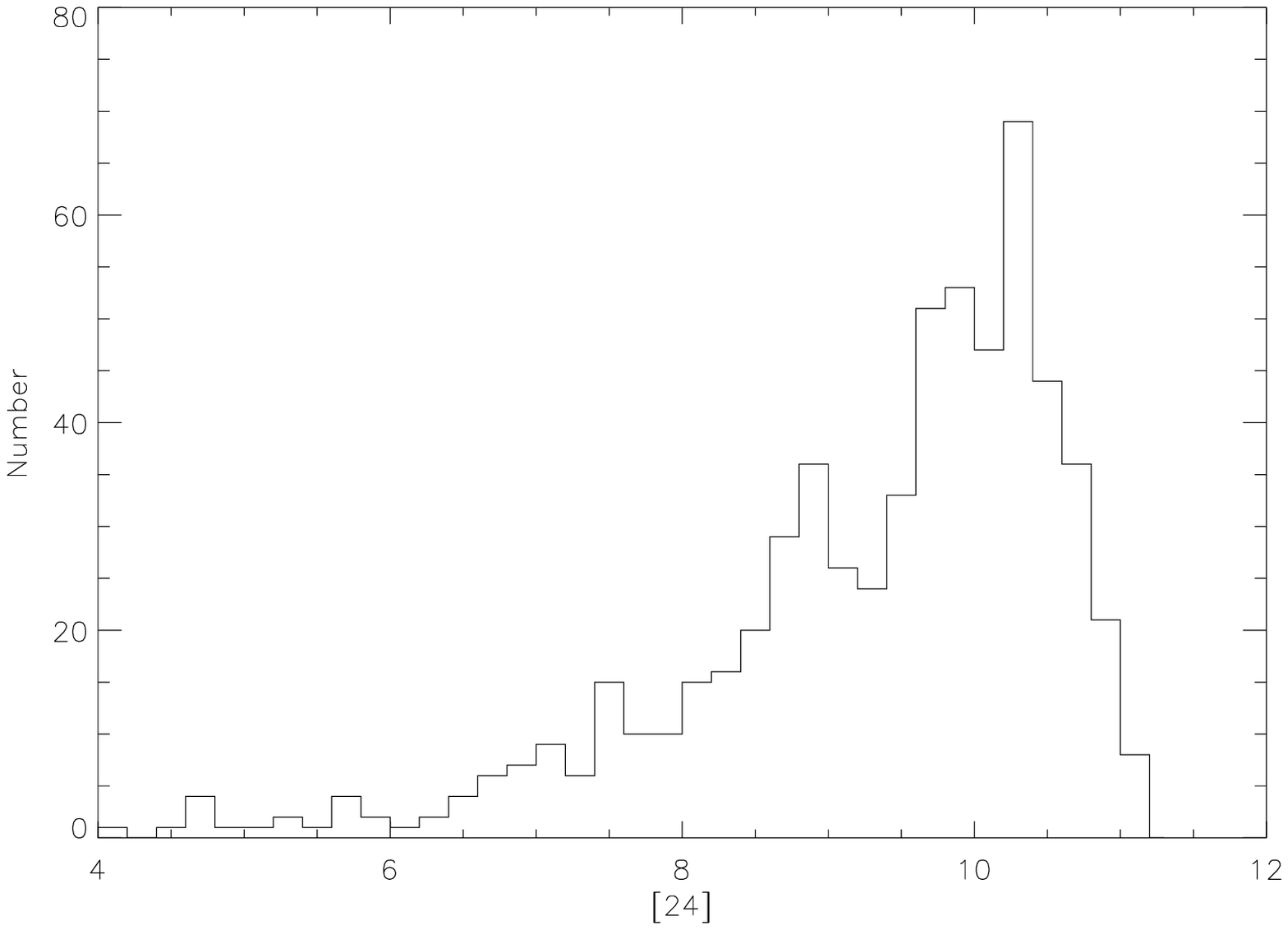}{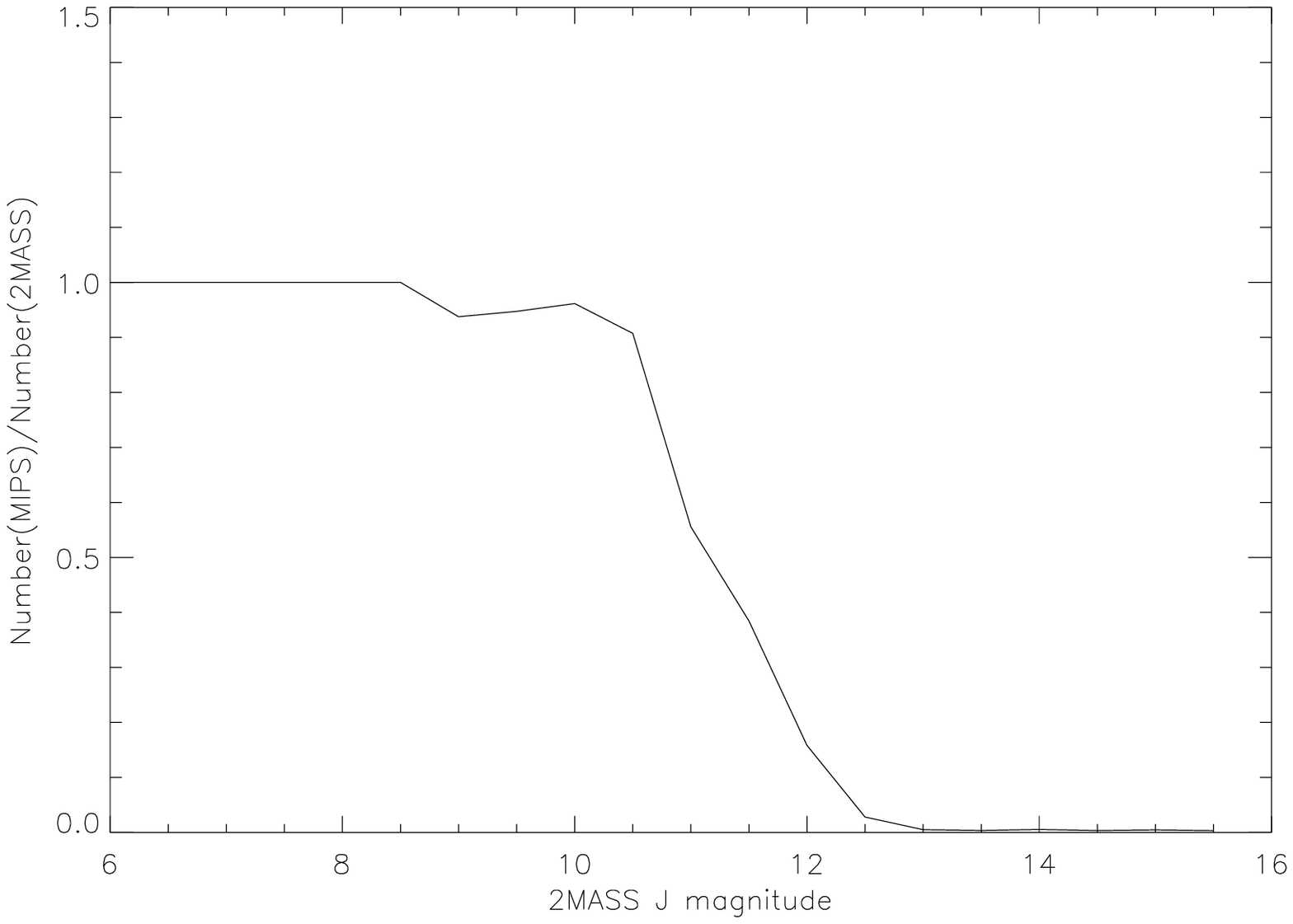}
\caption{(left) Distribution of [24] magnitudes
for sources detected at the 5$\sigma$ level with MIPS that have 2MASS/IRAC detections.  
The number counts peak at [24] $\sim$  10.5.  (right) The completeness profile as a function 
of 2MASS J magnitude.  Through J=10.5, $\gtrsim$ 90\% of all the 2MASS sources are detected 
with MIPS.  Only about half of the 2MASS sources between J=10.5-11 are detected with MIPS.}
\end{figure}
\begin{figure}
\plotone{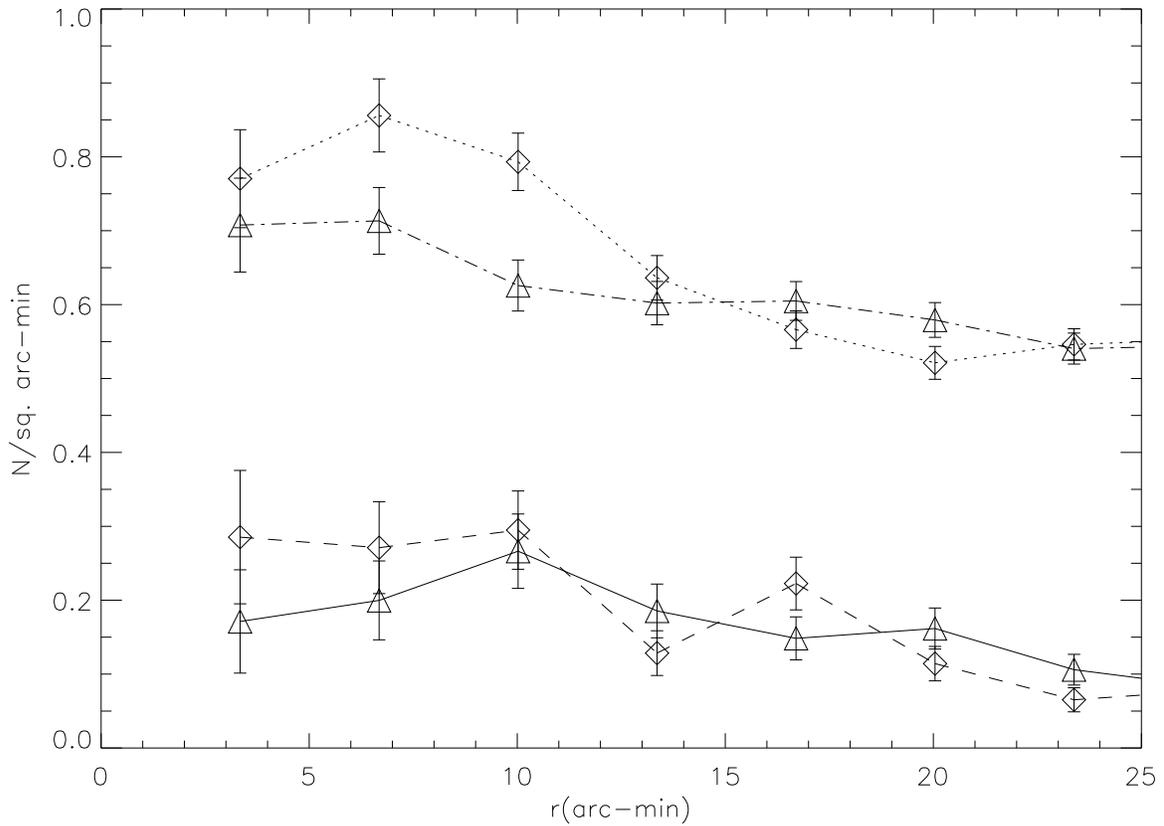}
\caption{The number density distribution of MIPS-detected sources as a function of distance from h Persei (diamonds; dashed line) and
$\chi$ Persei (triangles; solid line).  The error bars are based on Poisson statistics. The number counts decline through $\sim$ 20' away from
both cluster centers, though because of the small area coverage it is unclear exactly where the
true background lies.  For reference, we show the number density of 2MASS sources (J $\le$ 15.5) as a function of distance away from 
h Persei (diamonds; dotted line) and $\chi$ Persei (triangles; dot-dashed line) divided by 5.  The peak number density of MIPS sources is about an order of magnitude smaller than for
2MASS J band as found by C07a.} 
\label{dens}
\end{figure}

\begin{figure}
\plotone{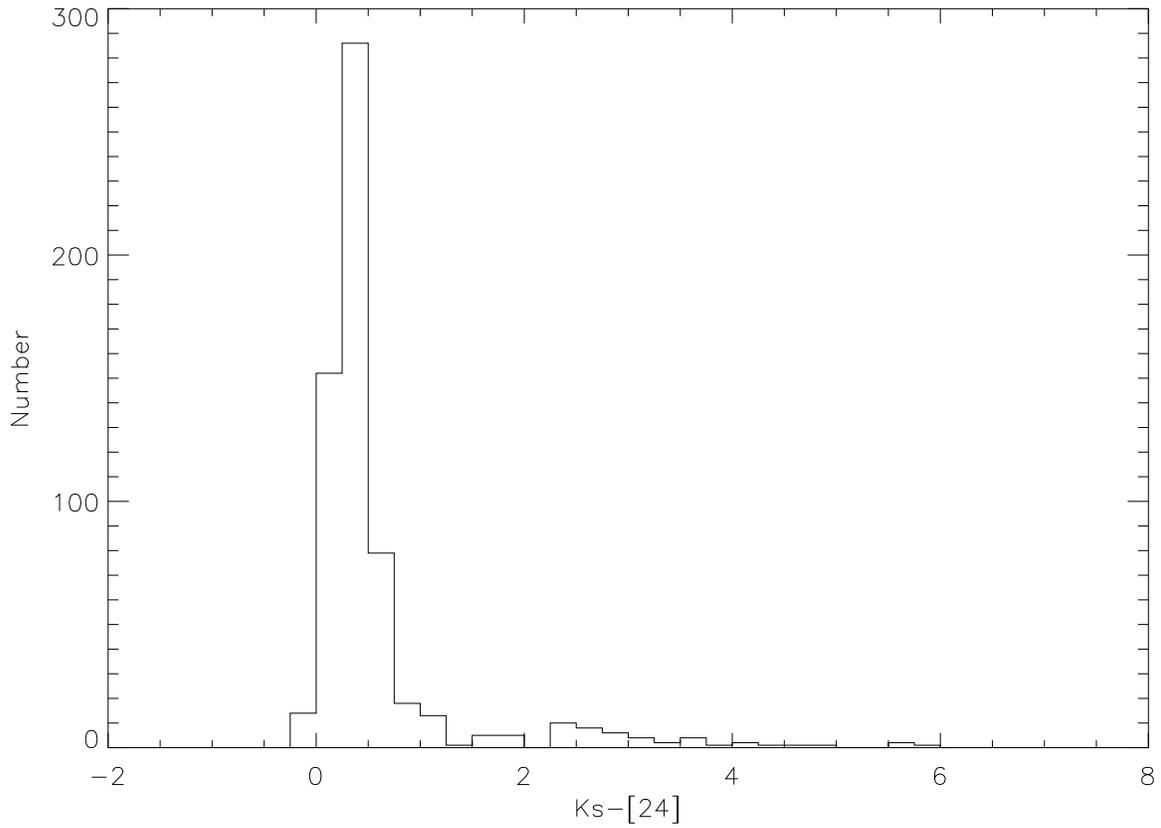}
\caption{Distribution of $K_{s}$-[24] colors for the 2MASS/MIPS detections.  The distribution
is peaked at $K_{s}$-[24] $\sim$ 0-0.5 and has a long positive tail stretching to $K_{s}$-[24] $\sim$ 6.}
\label{k24dist}
\end{figure}

\begin{figure}
\plotone{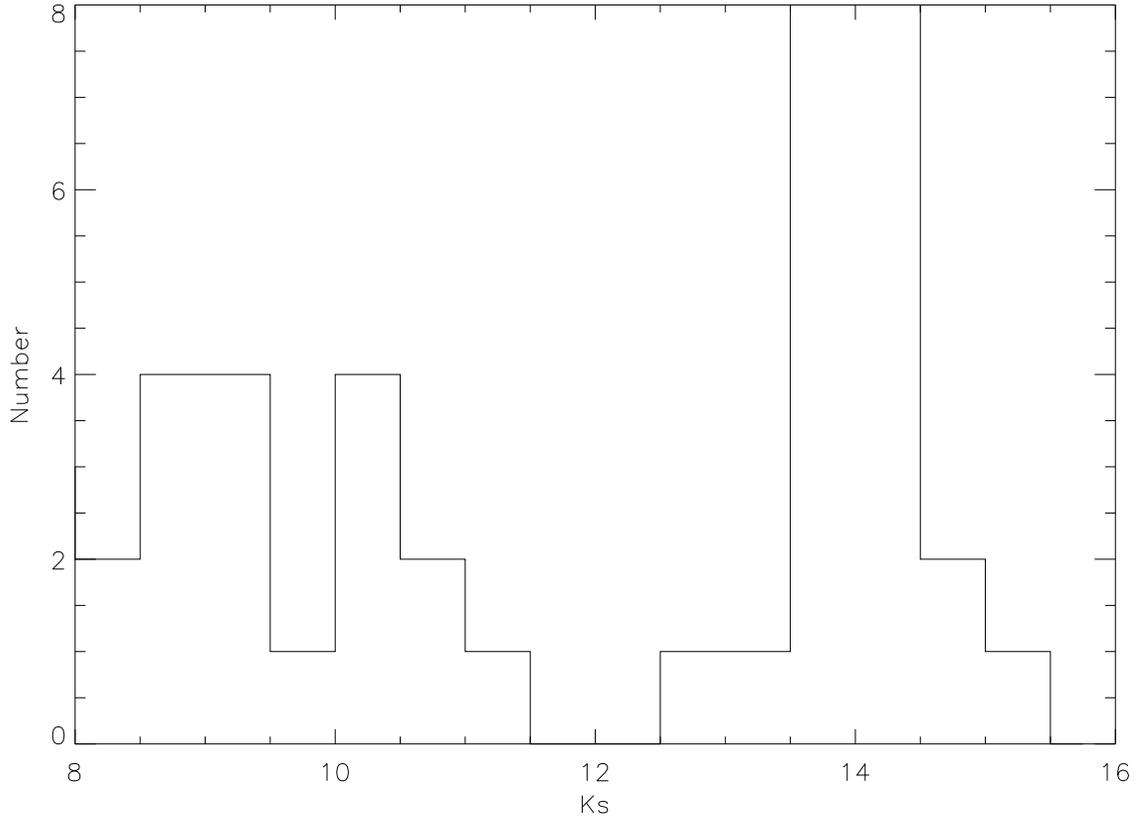}
\caption{Distribution of K$_{s}$ magnitudes with strong IR excess in the MIPS bands (K$_{s}$-[24]$\ge$2).
There appear to be two populations of strong MIPS excess sources: a bright distribution from 
K$_{s}$=9-11 and a faint distribution from K$_{s}$=13.5-15.}
\label{kexc}
\end{figure}

\begin{figure}
\plotone{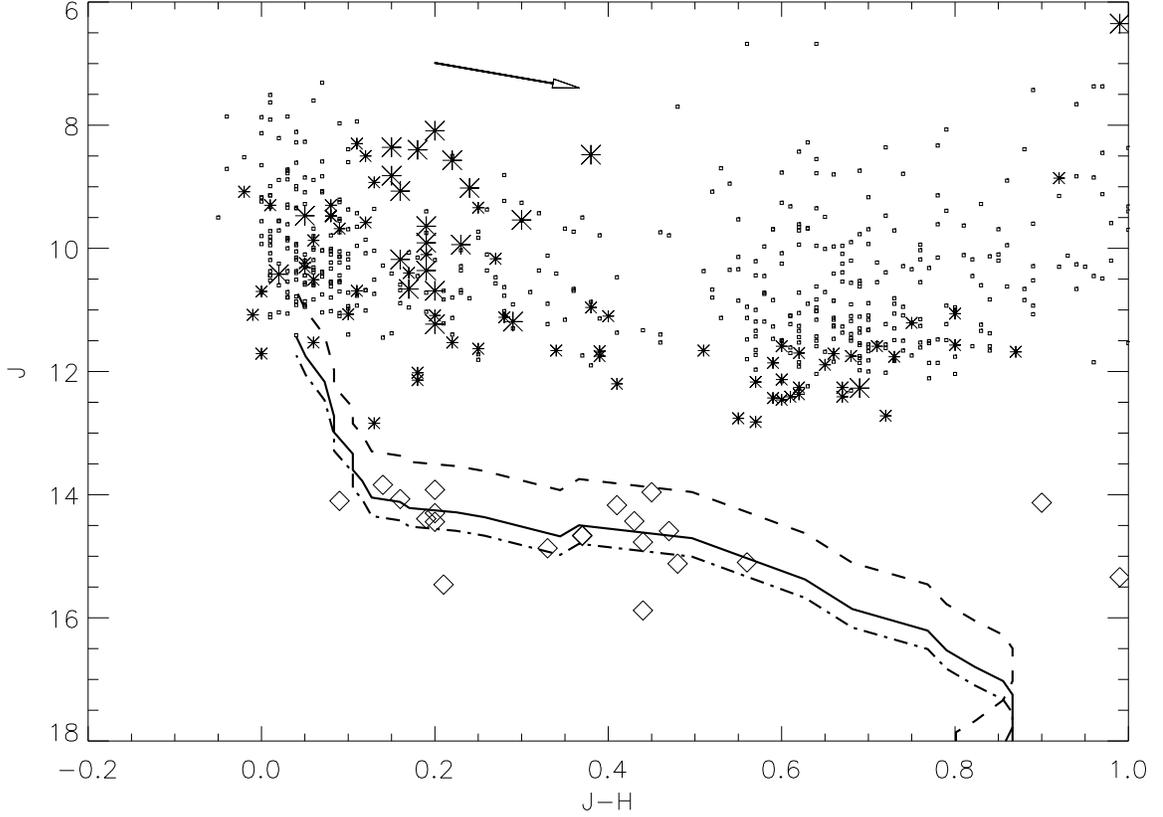}
\caption{Distribution of photospheric (small squares), marginal [24] excess (small asterisks for 
sources brighter than J=13), and 
strong [24] excess (large asterisks for J$\le$13; diamonds for J$\ge 13$) sources.  'Marginal' 
and 'strong' excesses are defined as sources 
with $K_{s}$-[24]=0.65-2 and $K_{s}$-[24]$\ge$2, respectively. The 14 Myr Siess et al. (2000) isochrone 
(solid line), reddened to A$_{V}$ = 1.62 (reddening vector shown as arrow), is plotted with a 0.3 magnitude lower bound (dot-dashed line) for photometric errors and a 0.75 magnitude upper bound (dashed line) for binarity.  There are two 
sources at J $\sim$ 14.66, J-H $\sim$ 0.36 that are not distinguishable on this plot.}  
\label{jjh}
\end{figure}
\begin{figure}
\plotone{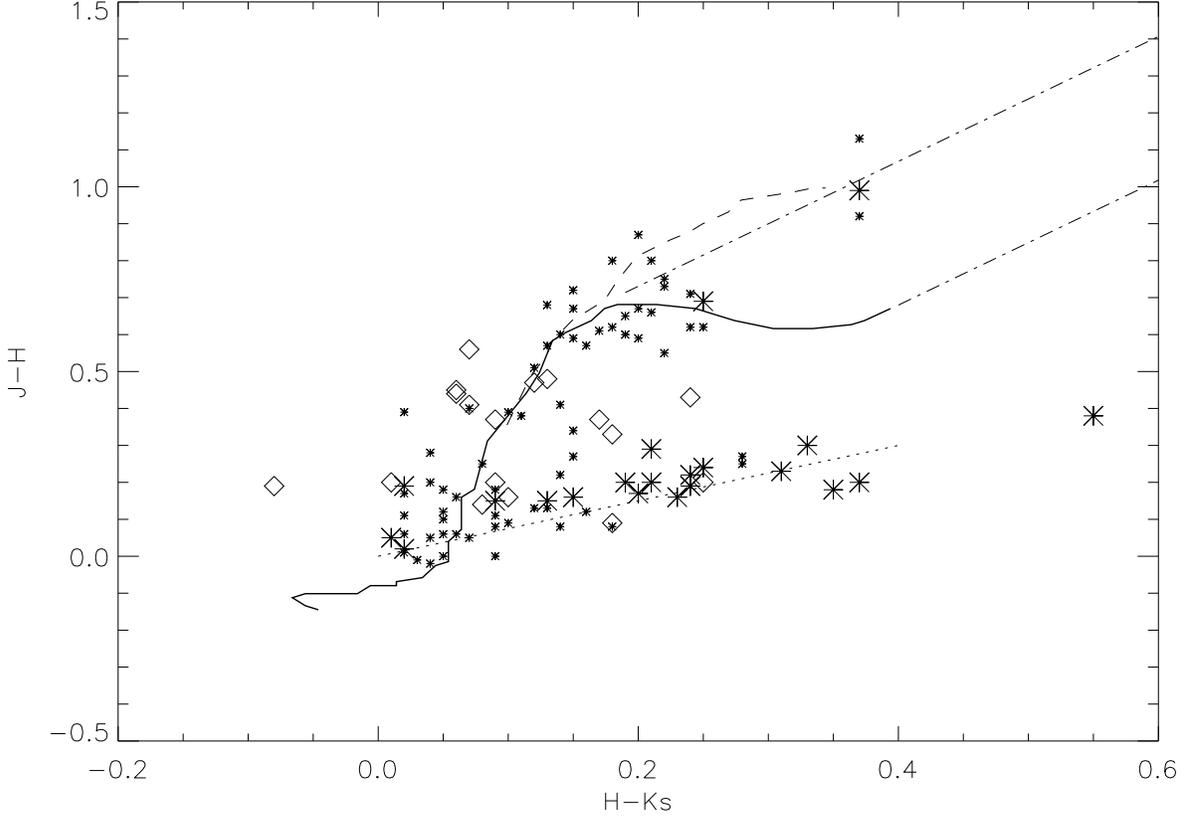}
\caption{Distribution of 24 $\mu m$ excess sources in observed J-H/H-K$_{s}$ color-color space.  The symbols are the same as in 
the previous figure.  The reddening band is overplotted as two straight dash-dotted lines.  
Only the faint sources lying along the 14 Myr isochrone are 
included. The bright sources with K$_{s}$ $\ge$ 2 follow a clear Be star locus (dotted line) from zero color 
to J-H=0.2, H-$K_{s}$=0.3; bright sources with weaker excess (K$_{s}$-[24] $\sim$ 0.65-2) appear either 
along the Be star locus or close to the giant locus (curved dashed line).  
The faint excess sources appear to be evenly distributed across the photospheric track (solid line).  
The IR-excess population is probably comprised of two 
main groups: bright Be stars with optically-thin free-free emission and fainter 
(J $\le$ 13.5) pre-main sequence A-F stars that likely harbor protoplanetary disks.} 
\end{figure}
\begin{figure}
\plottwo{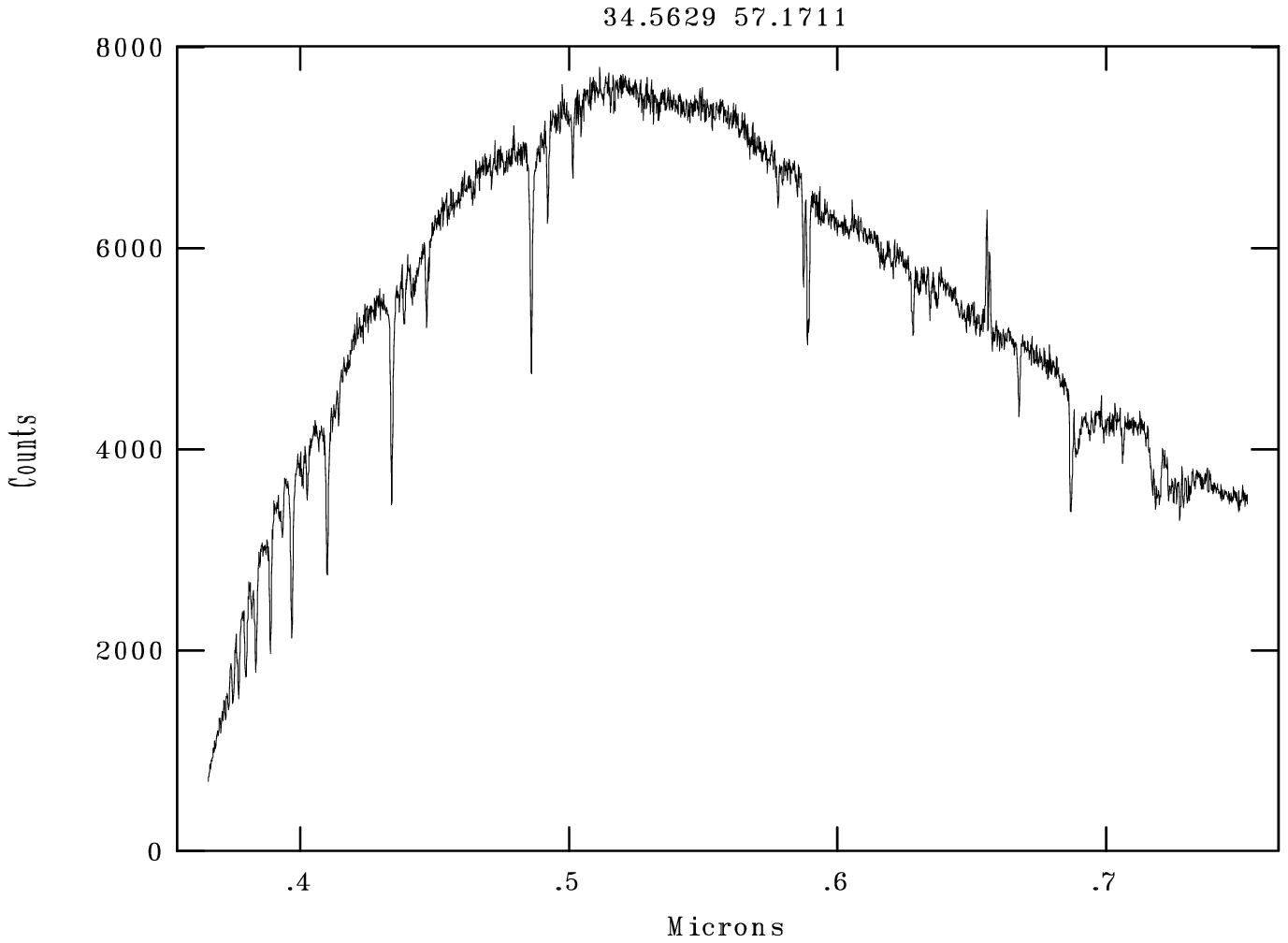}{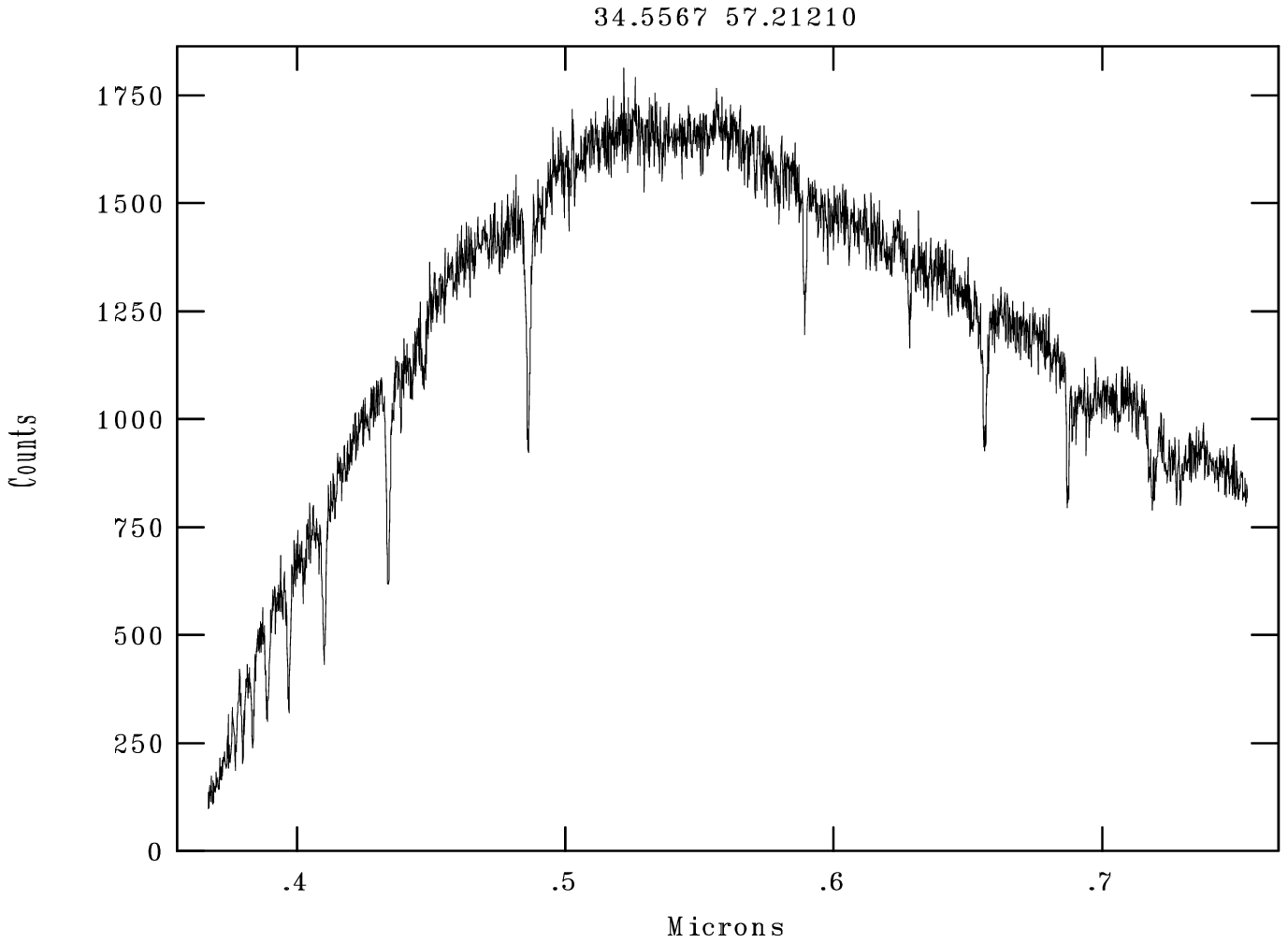}
\caption{Sample spectra of bright (J $\le$ 13) 24 $\mu m$ excess sources in the MIPS field.  The source on the left 
was identified as a Be star from Bragg \& Kenyon (2002) and has IR-excess emission from a circumstellar envelope.  
The source on the right has a spectral type of B4 and appears to be consistent with cluster membership.}
\end{figure}
\clearpage
\thispagestyle{empty}
\setlength{\voffset}{-18mm}
\begin{figure}
\plottwo{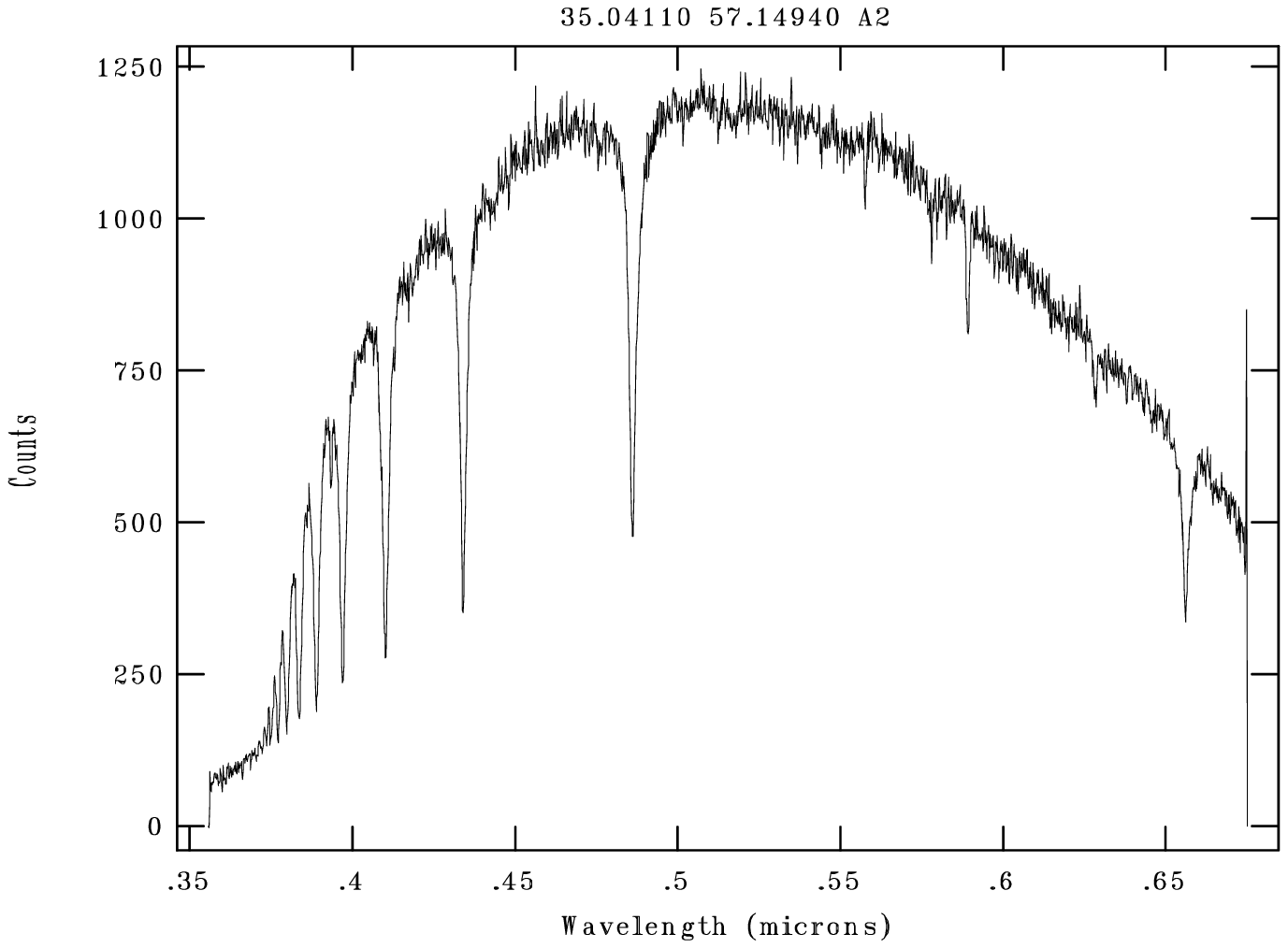}{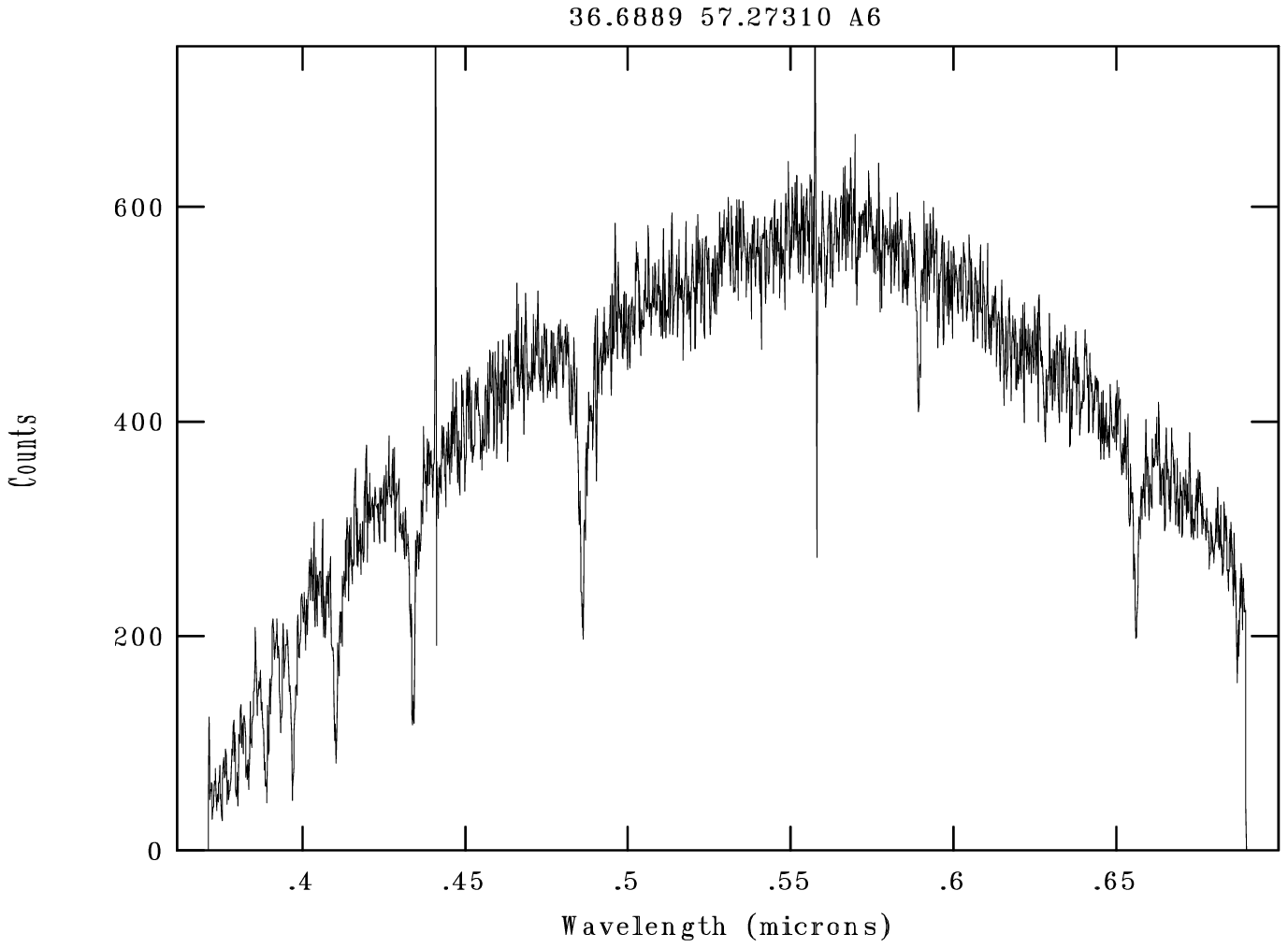}
\plottwo{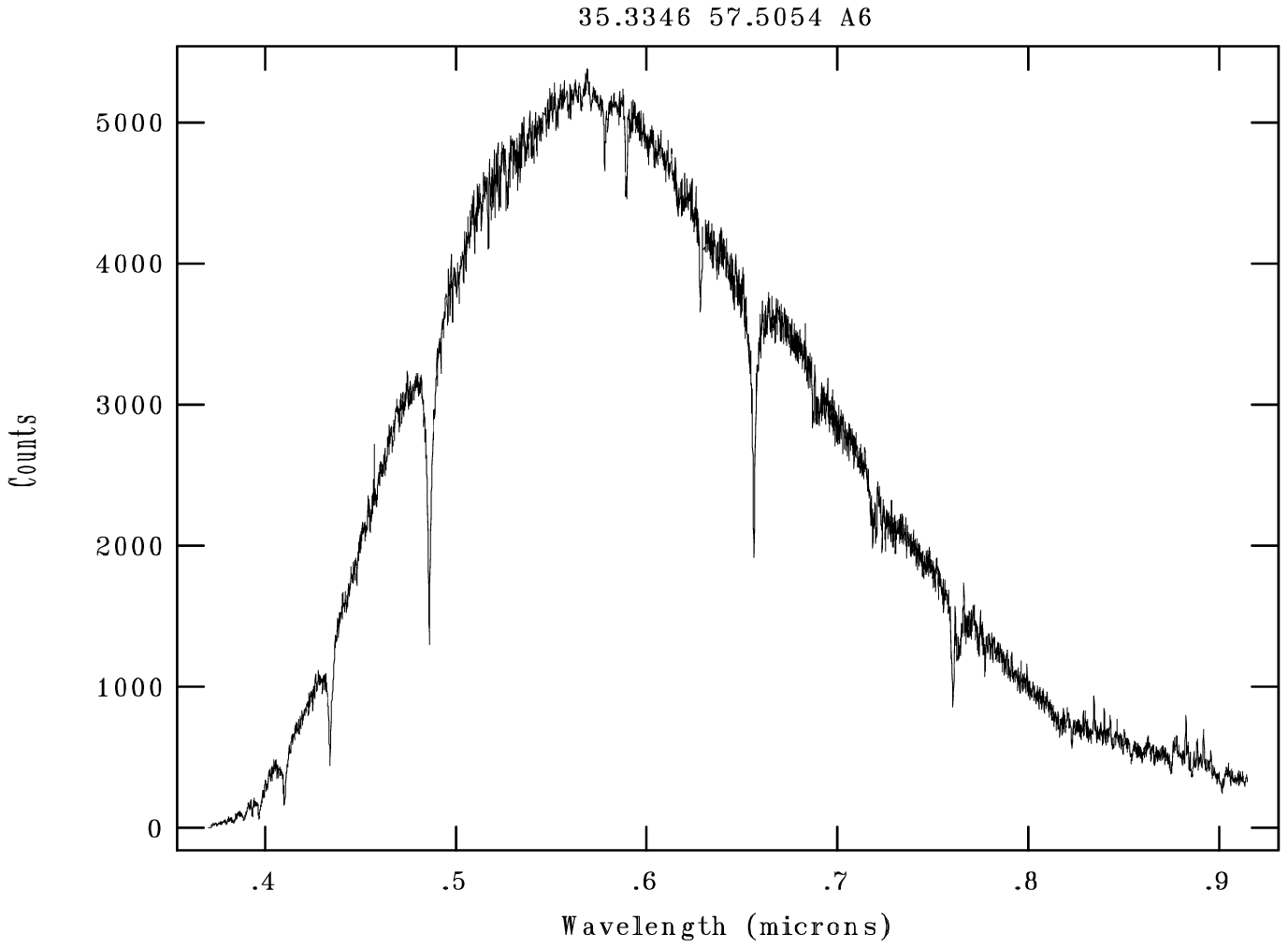}{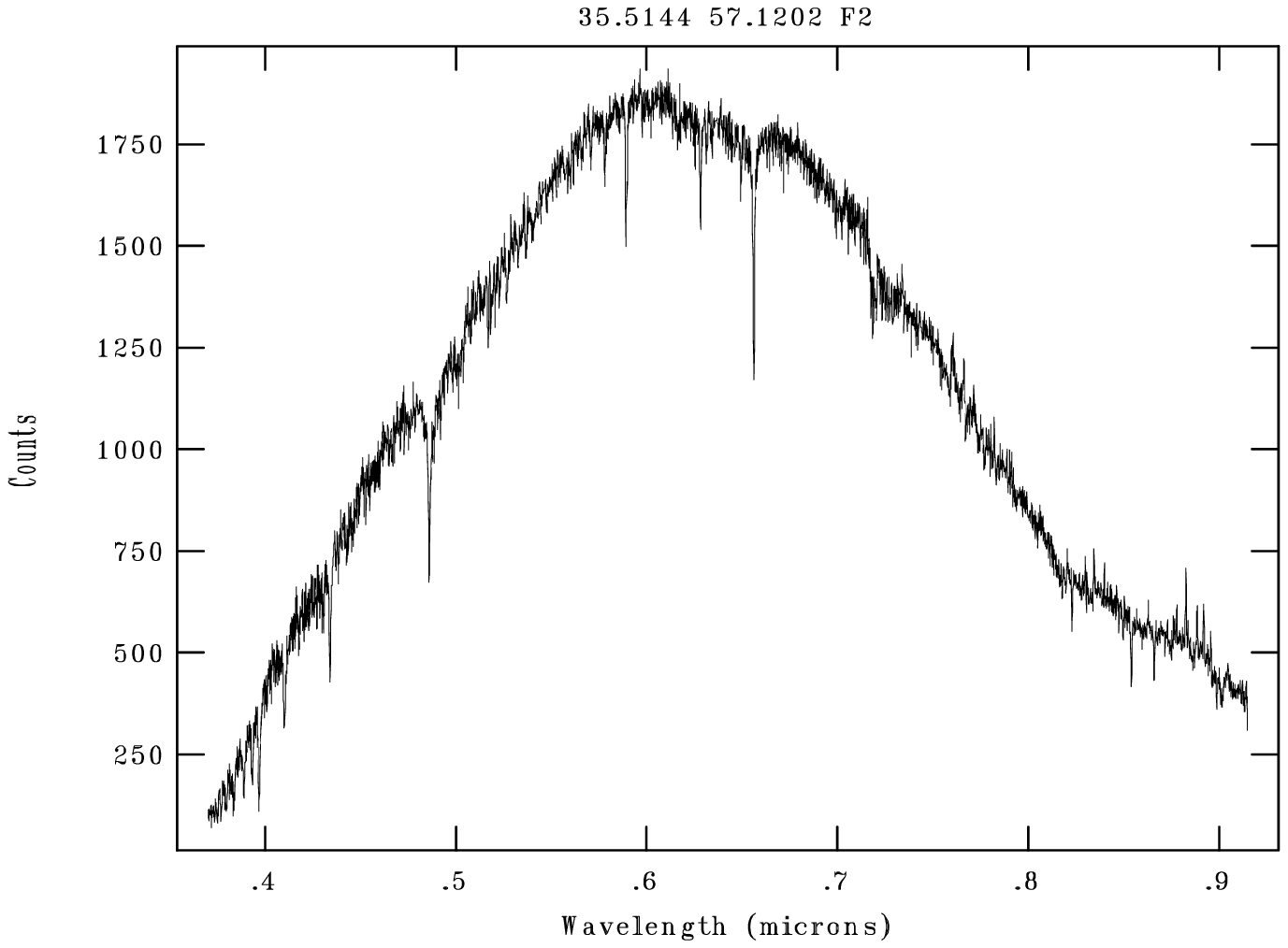}
\plottwo{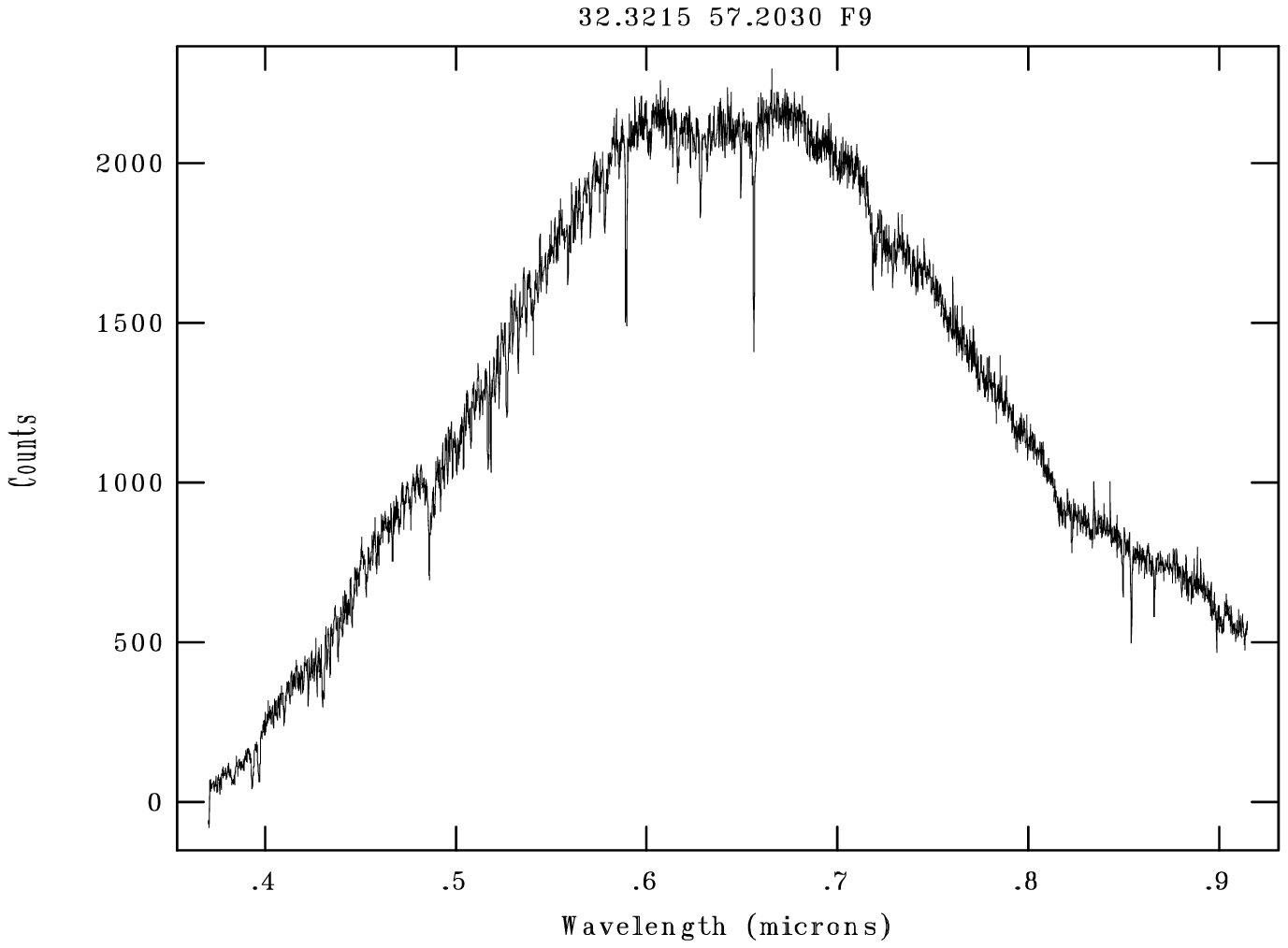}{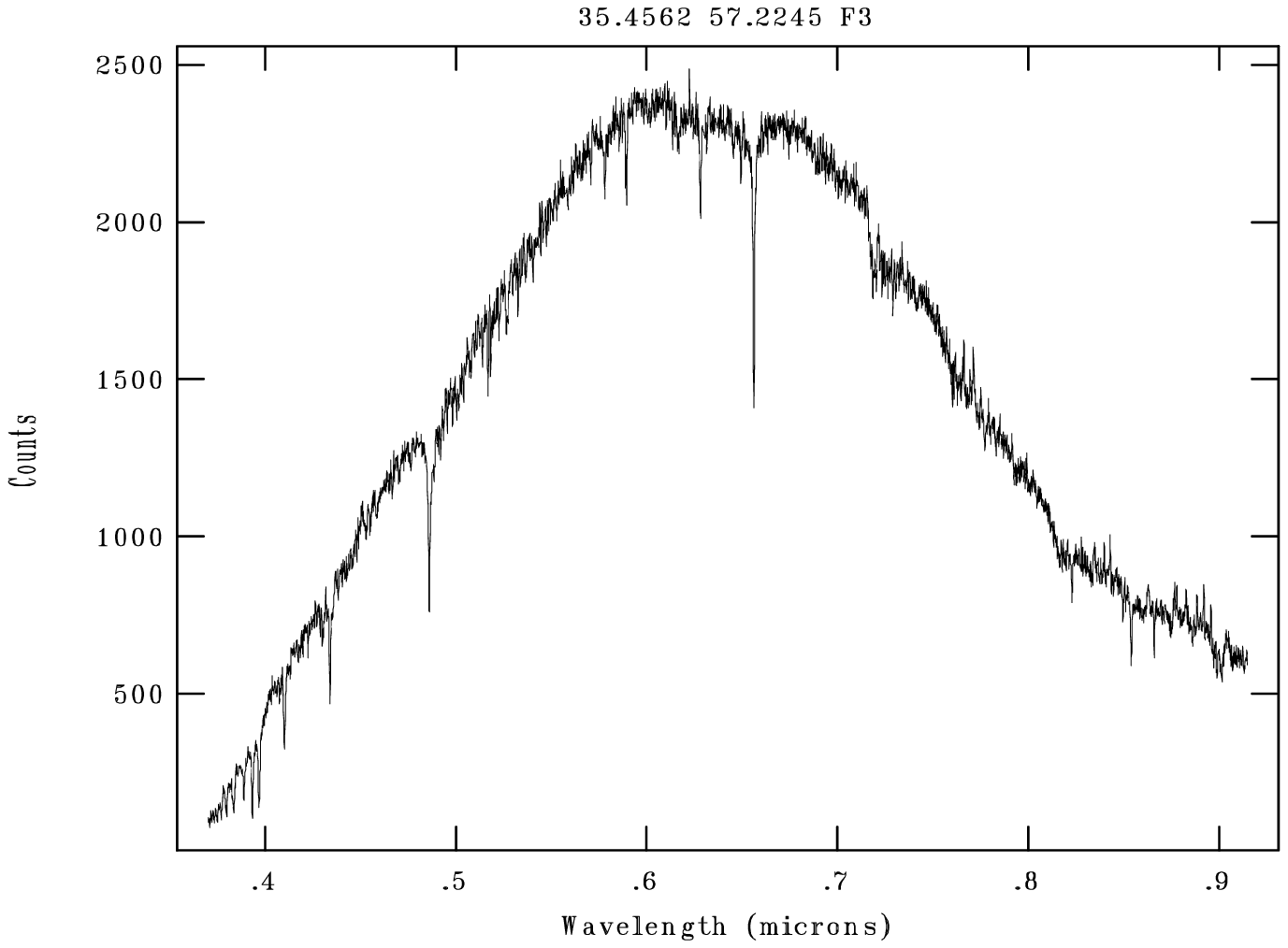}
\plottwo{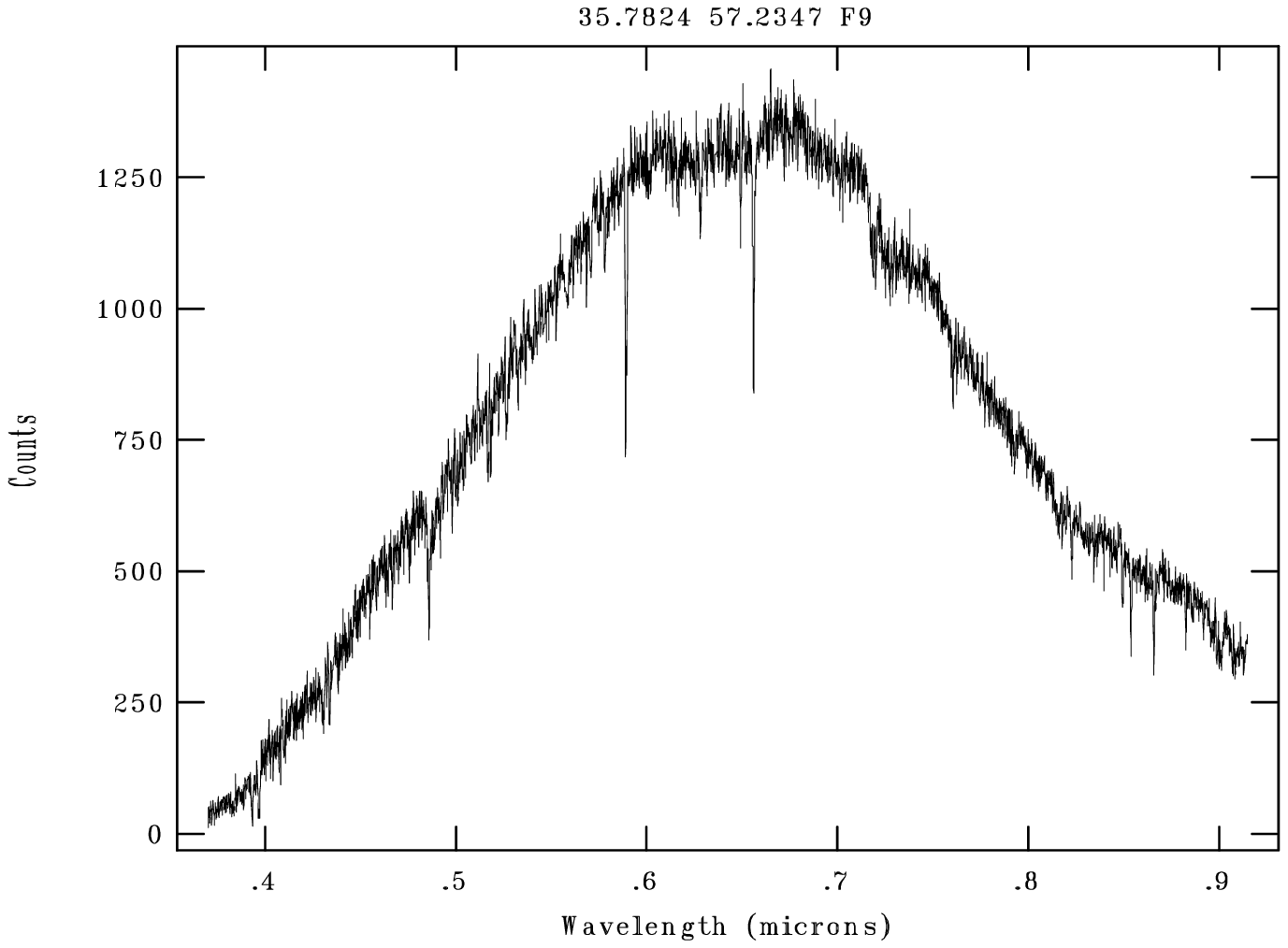}{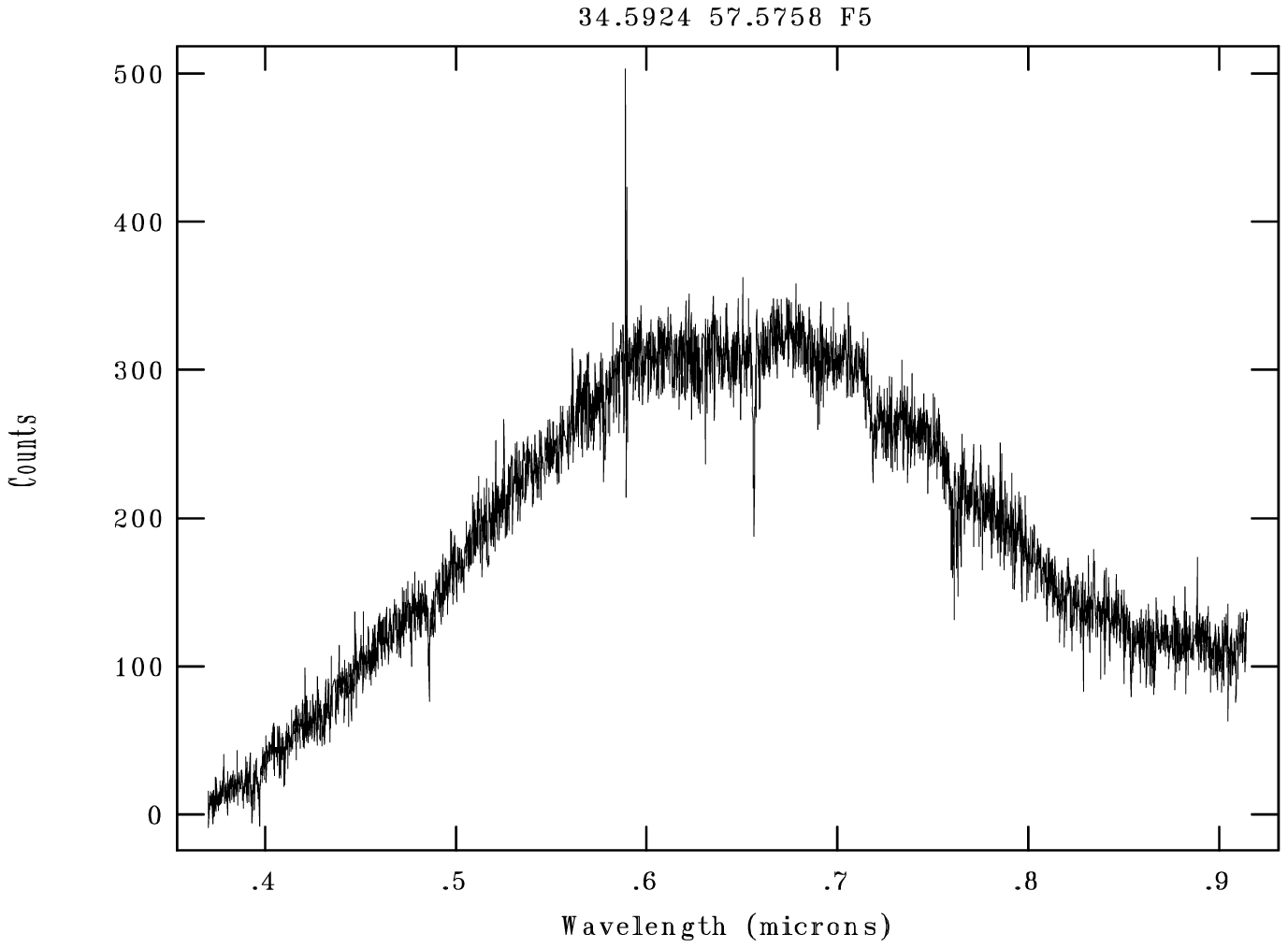}
\caption{Spectra for 8 faint (J $\ge$ 13) stars with 24 $\mu m$ excess observed with 
Hydra (top two) and Hectospec (bottom six).  The coordinates and spectral types for each 
source are listed on the top of each plot. 
The A6 star observed with Hydra has some image artifacts (spikes in count level well above the continuum level) at 
$\sim$ 0.44 and 0.56 \AA\ due to bad sky subtraction, though this did not impact the spectral type determination as 
we use other spectral indices that were not contaminated.  
The F5 star lies about one magnitude fainter than the 14 Myr isochrone 
for h and $\chi$ Persei (J$\sim$ 15.9) and is thus both photometrically and spectroscopically 
inconsistent with cluster membership.  The other stars are both photometrically and spectroscopically consistent 
with being $\sim$ 13-14 Myr old at a distance of 2.34 kpc.  
Thus, 7 of the 17 faint MIPS-excess sources 
lying on the J/J-H isochrone for h and $\chi$ Per are spectroscopically confirmed members.}
\end{figure}
\clearpage
\setlength{\voffset}{0mm}

\begin{figure}
\plotone{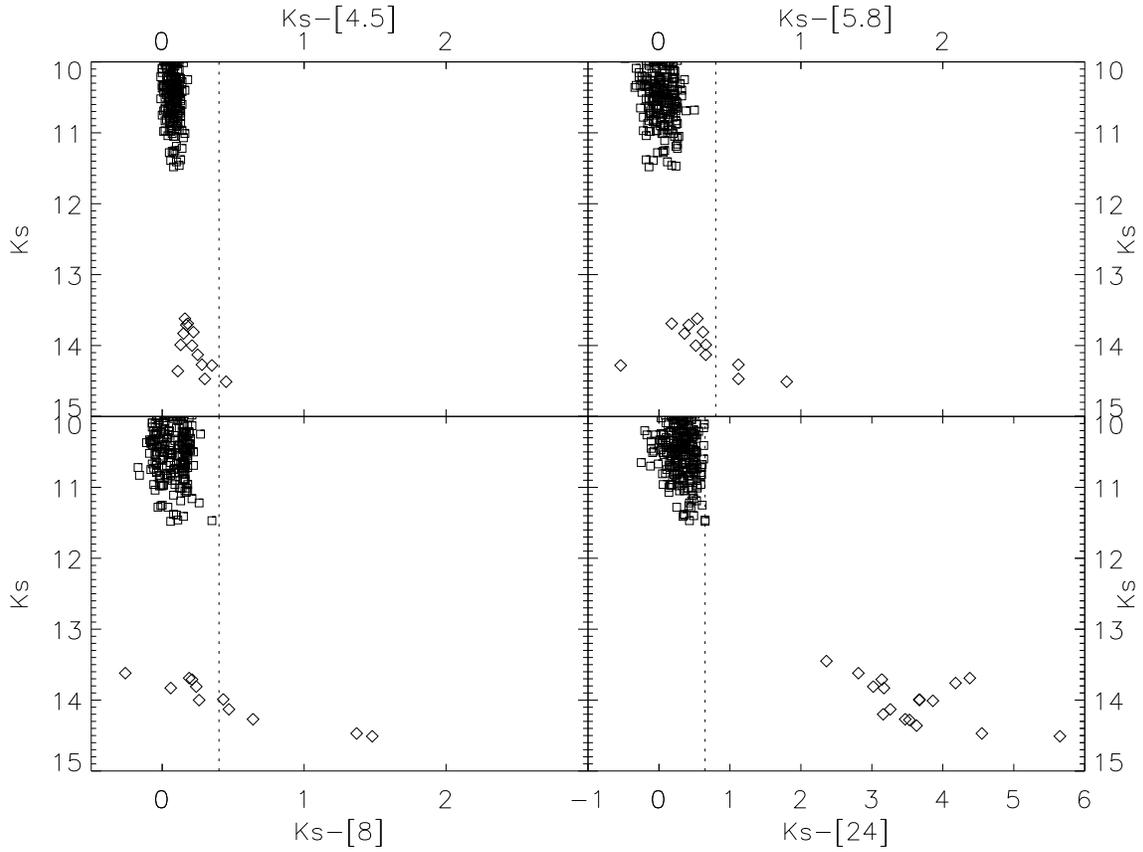}
\caption{K$_{s}$ vs. K$_{s}$-[4.5], K$_{s}$-[5.8], K$_{s}$-[8], and K$_{s}$-[24] 
color-magnitude diagrams for bright, photospheric sources and 24 $\mu m$ excess sources.  
Of the 24 $\mu m$ excess sources, 1/14 have excess at [4.5], 3/12 have excess at [5.8], and 
5/11 have excess at [8].} 
\label{cmdirac}
\end{figure}
\begin{figure}
\plotone{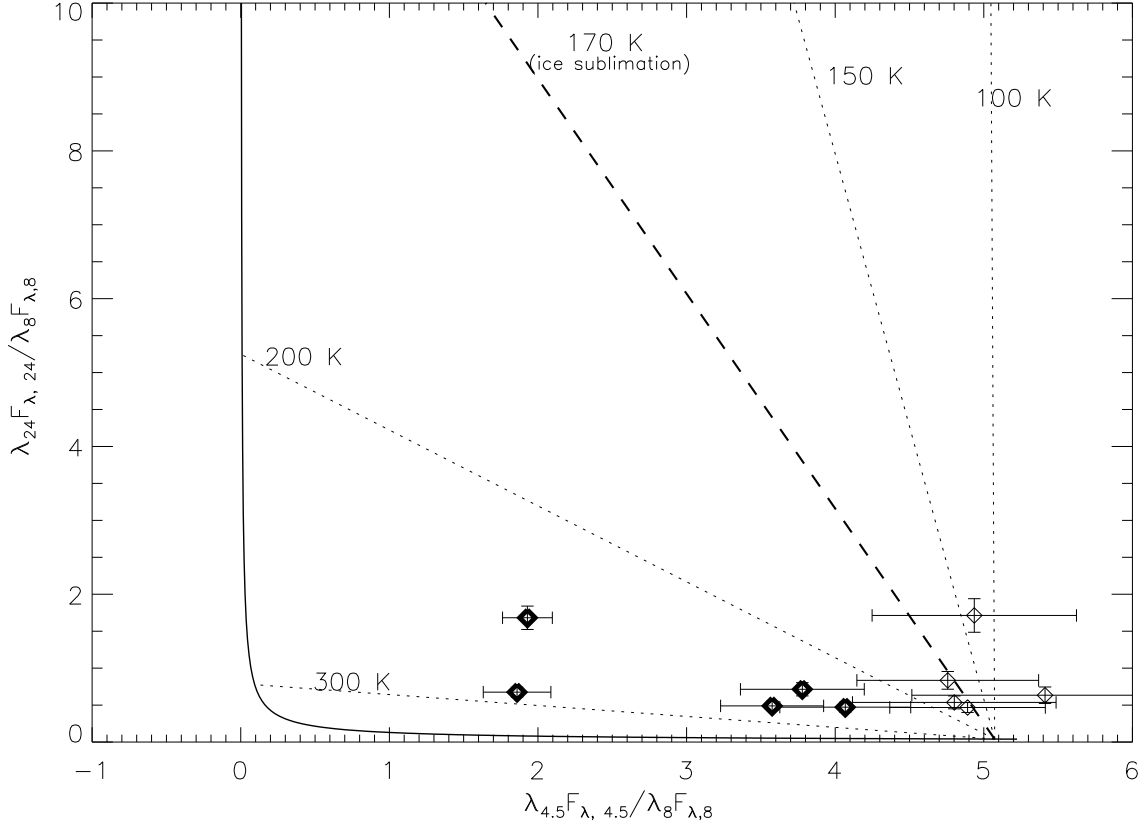}
\caption{Flux ratio diagram for sources with 5$\sigma$ detections at 4.5 through 24 $\mu m$ with 
errors in the flux ratios overplotted.
The flux ratios for single temperature blackbodies are shown as a solid line.  Dotted lines show 
loci of dust temperatures assuming a stellar blackbody temperature of 7250 K.  Sources 
with 8 $\mu m$ excess are shown in bold.  The five sources 
with 8 $\mu m$ excess typically have warmer dust temperatures (T $\sim$ 200-300 K) 
indicative of terrestrial zone emission.  The sources without 8 $\mu m$ excess have 
colder dust temperatures. 
Two sources probably have some dust at temperatures comparable 
to the water ice sublimation point (170 K).  Two sources probably have dust grains 
that are colder and icy ($\sim$ 100-150 K).  Typical errors for sources with only MIPS excess 
are $\sim$ 50 K, while errors for sources with IRAC and MIPS excess were smaller ($\sim$ 20 K).}
\label{fr}
\end{figure}

\clearpage
\thispagestyle{empty}
\setlength{\voffset}{-18mm}
\begin{figure}
\epsscale{0.85}
\plotone{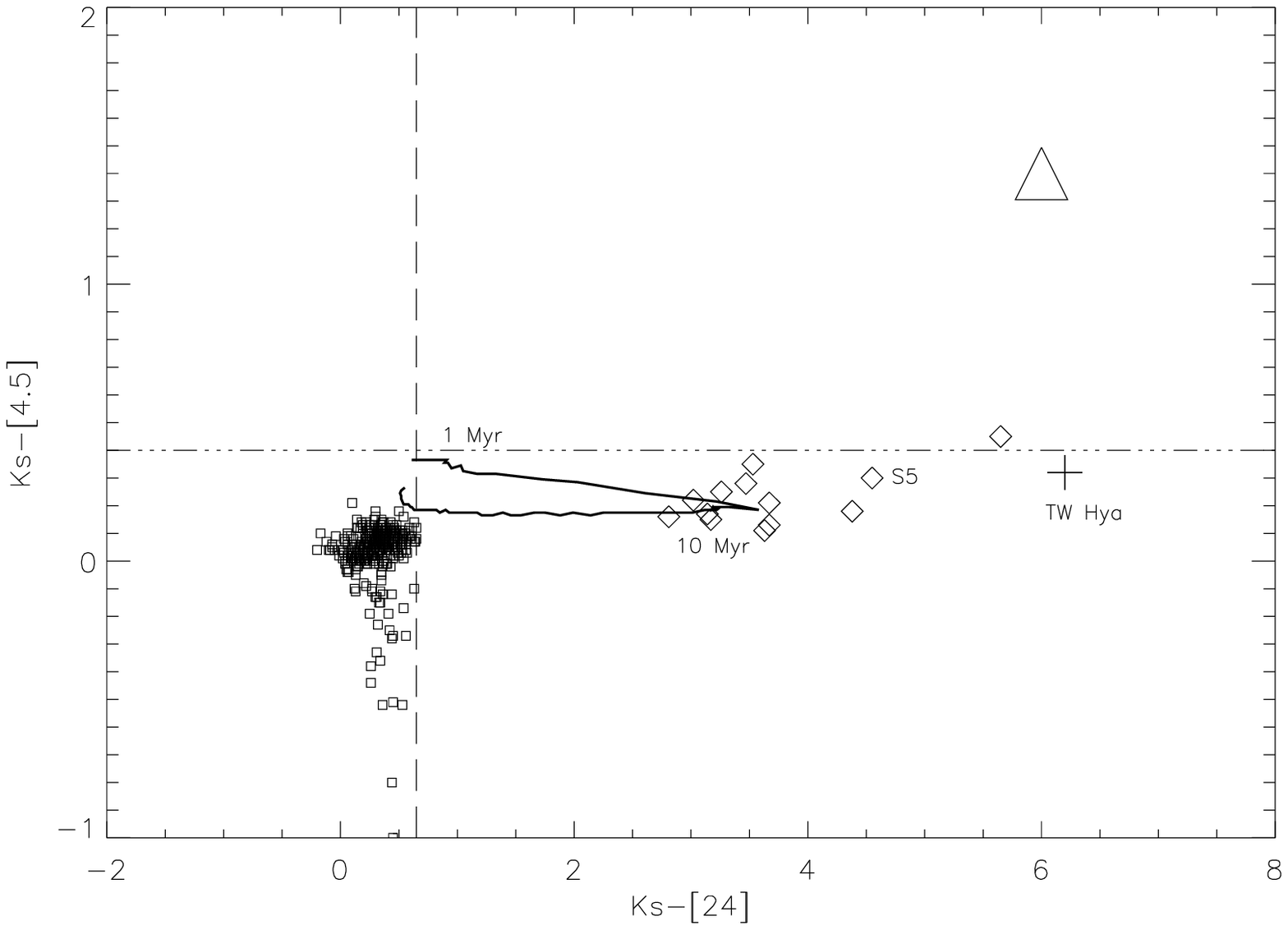}
\plotone{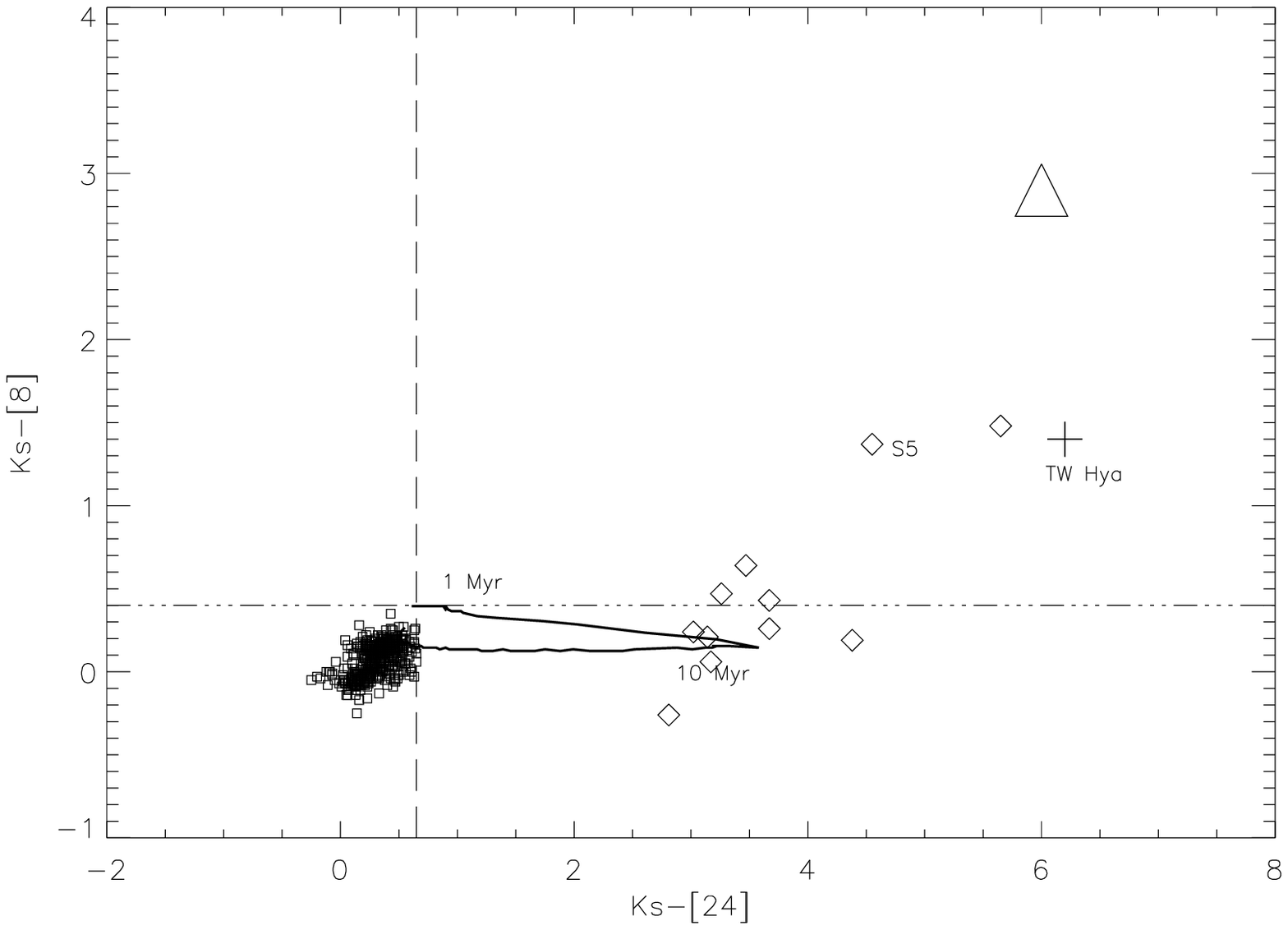}
\caption{The (top)$K_{s}$-[3.6]/$K_{s}$-[24] and (bottom) $K_{s}$-[8]/$K_{s}$-[24] color-color 
diagrams.  Sources below the horizontal line have photospheric [4.5] emission; those to the left of the 
vertical line have photospheric [24] emission.  The source with $K_{s}$-[8]$\sim$1.3 and $K_{s}$-[24]$\sim$4.5 
is source 5 in C07b and is labeled 'S5'.  One source appears to 
have mid-IR colors strongly resembling TW Hya (shown as a large cross; reddened to h and $\chi$ Per).  All sources have 
colors inconsistent with an optically-thick primordial disk (large triangle).  The debris disk 
locus is overplotted as a dark solid line with reddened K$_{s}$-[24] colors at 1 Myr ($\sim$ 0.9) and 10 Myr ($\sim$ 3.2) labeled. 
Debris from planet formation at 30-150 AU accurately reproduces the $K_{s}$-[4.5]/$K_{s}$-[24] diagram colors and 
is able to reproduce the $K_{s}$-[8] colors for sources with weaker [8] excess.}
\label{colcol}
\end{figure}
\clearpage
\setlength{\voffset}{0mm}

\begin{figure}
\plotone{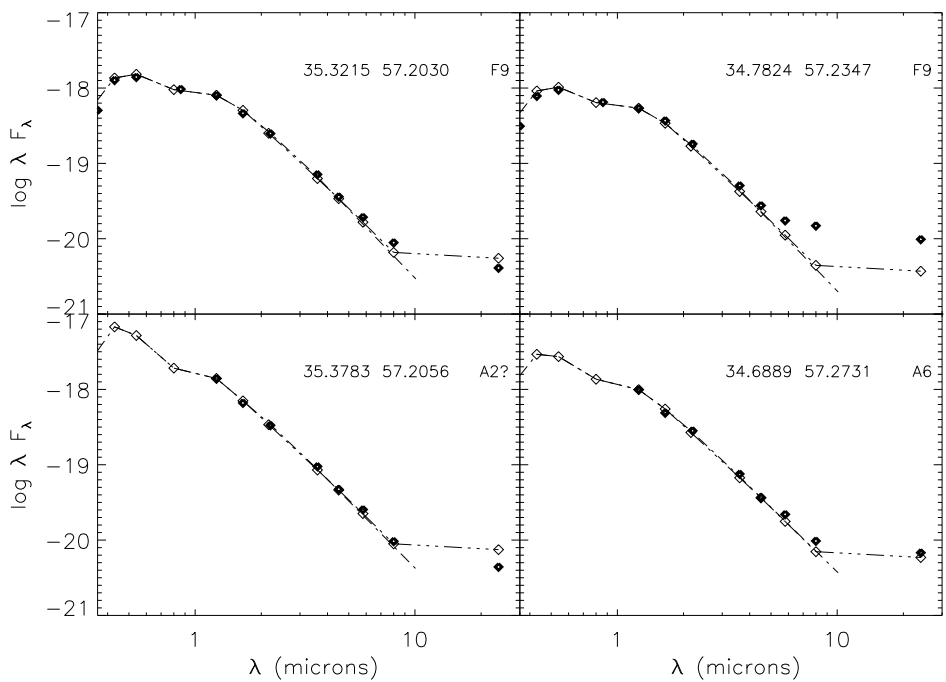}
\caption{Spectral energy distributions for selected MIPS-excess sources.  
The J2000 coordinates of the sources (in degrees) are 35.3215, 57.2030 (1); 34.7824, 57.2347 (2); 
34.6889, 57.2731 (3); and 35.3783, 57.2056 (4).  The spectral types of the 
sources are (clockwise from the top) F9, F9, A6, and A2.  The photospheric model (dash-dot) and 
cold debris disk model (dash-three dots) from Kenyon \& Bromley (2004) are overplotted.
The top-right F9 source was modeled as having terrestrial zone emission, though 
the disk may extend to more distant, cooler regions.  
Two sources (top-left F9 and A6) exhibit weak 8$\mu m$ emission, whereas the A2 source, typical 
of the majority of faint MIPS-excess sources, has no excess emission at 8$\mu m$.}
\label{sed4}
\end{figure}

\begin{figure}
\epsscale{0.75}
\plotone{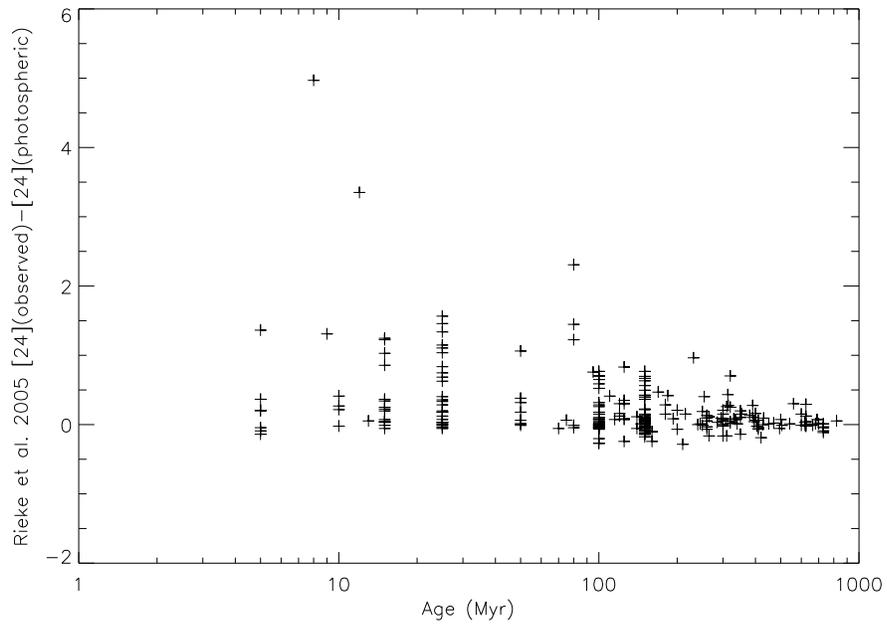}
\caption{(top) Color-excess of sources from the R05 sample assuming that the photospheric $K_{s}$-[24] color 
is $\sim$ 0.  The strongest excess sources in this sample are HR 4796A (8 Myr) and $\beta$ Pictoris (12 Myr).  
There is a clear trend of decreasing color excess vs. age beyond $\sim$ 20 Myr.} 
\label{excvagegr}
\end{figure}
\clearpage
\begin{figure}
\plotone{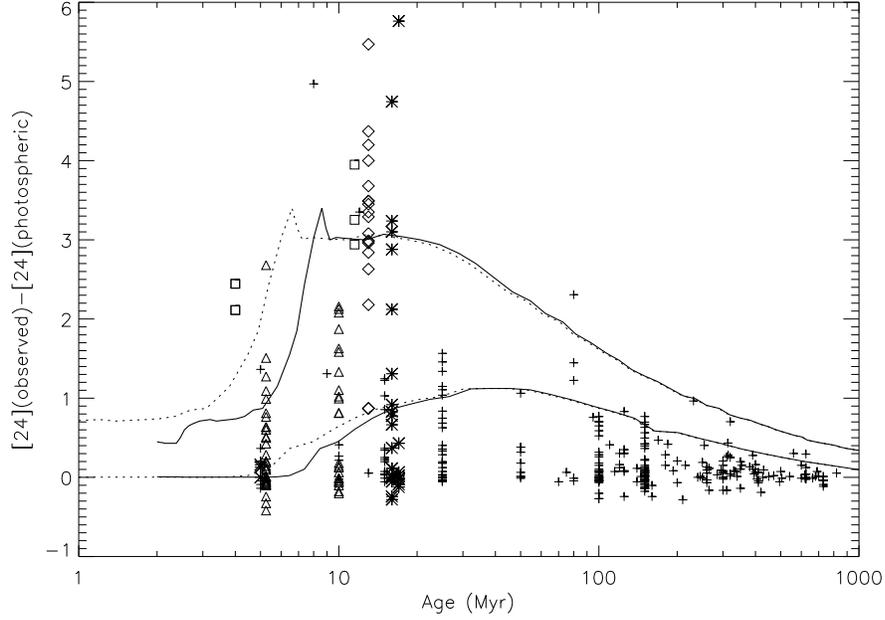}
\caption{Color-excess of sources from R05 as well as $\sim$ 13 Myr-old h and $\chi$ Persei (diamonds);
Sco-Cen subgroups at 5, 16, and 17 Myr old (asterisks; Chen et al. 2005b); 4 and 
11.8 Myr-old Cepheus subgroups Tr 37 and NGC 7160 (squares; Sicilia-Aguilar et al. 2006); and 5 and 10 Myr-old 
Orion Ob1b and Ob1a (Hernandez et al. 2006).  We include the B4 star in h and $\chi$ Per with 
MIPS excess.  HR 4796A is again clearly visible while $\beta$ Pic is obscured by the Cepheus data.
Upper Sco data (5 Myr old) is also obscured by Orion Ob1b and has 24$\mu m$ excess $\sim$ 0-0.3.
 Overplotted are debris disk evolution tracks from 
Kenyon \& Bromley (2004) for a 3$\times$ and 1/3$\times$ (scaled) MMSN disk assuming that primordial disk 
grains grow to planetesimal sizes by 0 (dotted line) and 2 Myr (solid line).  There is a clear trend of increasing 
color excess at 24 $\mu m$ from $\sim$ 5-10 Myr, a peak at $\sim$ 10-15 Myr, and a steady decline 
afterwards through 1 Gyr.  The trend is clear even if the strongest excess sources in 
h and $\chi$ Per and Sco-Cen ([24]$_{obs}$ -[24]$_{\star}$ $\gtrsim$ 4-5) were excluded due to their uncertain evolutionary states.}
\label{excvage}
\end{figure}

\begin{figure}
\plotone{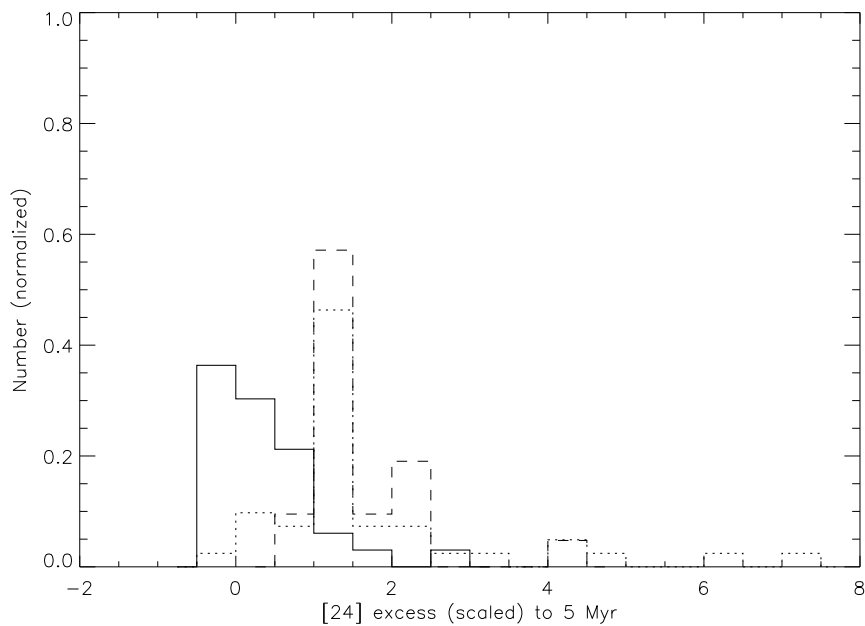}
\caption{[24]-[24]$_{\star}$ excesses for Orion Ob1b (solid line) compared to the 'scaled' 5 Myr Sco-Cen (dotted line) 
and Rieke et al. colors (dashed line), assuming that 
excess emission declines as t$^{-1}$ for t$\gtrsim$ 5 Myr.  We normalize the total number of sources in each bin by the total 
number of sources in all bins.  The excesses from many Sco-Cen and Rieke et al. sources are 
far redder than any Orion sources ($\gtrsim$ 4 magnitudes).  Thus, the evolution of 24 $\mu m$ emission from disks is not consistent 
with a t$^{-1}$ power law.  Further statistical tests, described in \S 4.2, verify that the emission rises from 5 to 10 Myr.}
\label{scaled}
\end{figure}
\end{document}